\documentclass[%
reprint,
superscriptaddress,
amsmath,amssymb,
aps,
prb,
]{revtex4-2}

\usepackage[utf8]{inputenc} \usepackage[T1]{fontenc}
\usepackage{physics}
\usepackage[colorlinks=true]{hyperref}
\usepackage{graphicx}
\usepackage{algorithm}
\usepackage{algpseudocode}
\usepackage{setspace}
\usepackage{physics}

\renewcommand{\vec}[1]{\mathbf{#1}}

\newcommand{\Eloc}{E_{\textrm{loc}}}

\newcommand{\pder}[2]{\frac{\partial #1}{\partial #2}}
\newcommand{\mean}{\overline{\sum}}
\newcommand{\Epsilon}{\mathcal{E}}
\newcommand{\dt}{\delta t}
\newcommand{\variance}[2]{\ensuremath{\textrm{Var}_{#2}\left(#1\right)}}

\newcommand{\changed}[1]{#1}

\begin{document}

\title{Ab-initio variational wave functions for the time-dependent many-electron Schr\"odinger equation}

\author{Jannes Nys}
\affiliation{Institute of Physics, École Polytechnique Fédérale de Lausanne (EPFL), CH-1015 Lausanne, Switzerland}
\affiliation{%
Center for Quantum Science and Engineering, \'{E}cole Polytechnique F\'{e}d\'{e}rale de Lausanne (EPFL), CH-1015 Lausanne, Switzerland
}

\author{Gabriel Pescia}
\affiliation{Institute of Physics, École Polytechnique Fédérale de Lausanne (EPFL), CH-1015 Lausanne, Switzerland}
\affiliation{%
Center for Quantum Science and Engineering, \'{E}cole Polytechnique F\'{e}d\'{e}rale de Lausanne (EPFL), CH-1015 Lausanne, Switzerland
}

\author{Alessandro Sinibaldi}
\affiliation{Institute of Physics, École Polytechnique Fédérale de Lausanne (EPFL), CH-1015 Lausanne, Switzerland}
\affiliation{%
Center for Quantum Science and Engineering, \'{E}cole Polytechnique F\'{e}d\'{e}rale de Lausanne (EPFL), CH-1015 Lausanne, Switzerland
}

\author{Giuseppe Carleo}
\email{giuseppe.carleo@epfl.ch}
\affiliation{Institute of Physics, École Polytechnique Fédérale de Lausanne (EPFL), CH-1015 Lausanne, Switzerland}
\affiliation{%
Center for Quantum Science and Engineering, \'{E}cole Polytechnique F\'{e}d\'{e}rale de Lausanne (EPFL), CH-1015 Lausanne, Switzerland
}

\begin{abstract}
Understanding the real-time evolution of many-electron quantum systems is essential for studying dynamical properties in condensed matter, quantum chemistry, and complex materials, yet it poses a significant theoretical and computational challenge.  Our work introduces a variational approach for fermionic time-dependent wave functions, surpassing mean-field approximations by accurately capturing many-body correlations. Therefore, we employ time-dependent Jastrow factors and backflow transformations, which are enhanced through neural networks parameterizations. To compute the optimal time-dependent parameters, we utilize the time-dependent variational Monte Carlo technique and a new method based on Taylor-root expansions of the propagator, enhancing the accuracy of our simulations. The approach is demonstrated in three distinct systems. In all cases, we show clear signatures of many-body correlations in the dynamics. The results showcase the ability of our variational approach to accurately capture the time evolution, providing insight into the quantum dynamics of interacting electronic systems, beyond the capabilities of mean-field.
\end{abstract}

\date{\today}

\maketitle

\section{Introduction}
Understanding the consequences of interactions in quantum many-electron systems is necessary for a wide range of applications, including the prediction of the electronic structure in quantum chemistry~\cite{choo2020fermionic,bauer2020quantum}, properties of crystalline solids~\cite{foulkes2001quantum}, insulating or (super)conducting behavior of complex materials~\cite{arovas2022hubbard, savary2016quantum}. In addition to experiments, our main source of available information originates from numerical simulations of these systems. A significant amount of effort has been devoted to developing scalable and accurate techniques to obtain the (equilibrium) ground and excited states of interacting many-body systems, by solving the time-independent Schr\"odinger equation (TISE)
\begin{align}
    \hat{H}\ket{\Psi} = E \ket{\Psi} . \label{eq:tise}
\end{align}

Real-time electron dynamics encompasses the explicit consideration of the time evolution of a quantum electronic system out of equilibrium. 
The problem boils down to capturing the time-dependent quantum state $\ket{\Psi(t)}$, by solving the time-dependent Schr\"odinger equation (TDSE), 
\begin{align}
    i\dv{\ket{\Psi(t)}}{t} = \hat{H}(t) \ket{\Psi(t)} , \label{eq:tdse}
\end{align}
subject to a time-dependent Hamiltonian $\hat{H}(t)$. Exploration of the time-dependent characteristics of quantum systems extends across various disciplines, encompassing phenomena such as correlated electrons in metal clusters, quantum dots, and ultracold Fermi gases~\cite{baletto2005structural, filinov2001wigner, filinov2000path, reimann2002electronic, giorgini2008theory}. Within these systems, there exists a rich interplay of static and dynamic behaviors, often governed by collective modes arising from strong coupling effects~\cite{moritz2003exciting, giorgini2008theory, bauch2010quantum}.

Real-time quantum dynamics has received comparatively less attention than solving the TISE, primarily because of its significantly increased difficulty level. \changed{The latter is a consequence of the fact that a quantum system tends to traverse a larger portion of the configuration (Hilbert) space than the subspace containing ground states. Therefore, heuristics developed for static problems are not directly applicable. More specifically, it is no longer feasible to restrict the calculations to the low-entanglement corner of the Hilbert space~\cite{eisert2015qmboutofequi, eisert2008area, eisert2006general}.
} 
Obtaining the time-evolving quantum state of a time-dependent Hamiltonian, as described by the TDSE in Eq.~\eqref{eq:tdse}, is traditionally achieved through a series of approximations.
A commonly adopted approach to solving the TDSE approximately is to ignore electron-electron correlations, as is done in mean-field approximations.
The first work in this direction comes from Dirac and Frenkel, from whom the time-dependent Hartree-Fock (TDHF) equations were obtained for electronic systems~\cite{dirac_note_1930, frenkel1934wave}. Later, MacLachlan improved on this method~\cite{mclachlan1964time}.
Furthermore, TDHF has been widely used in nuclear physics to study reactions~\cite{koonin1977time, li2005time}. However, TDHF underestimates the particle correlations, making it a suitable method for describing weakly correlated systems, such as many closed-shell molecules, but unreliable for strongly correlated ones, such as molecules with significant electron-electron interactions. To build correlations on top of TDHF, this framework was extended to the multi-configuration time-dependent Hartree-Fock method (MC-TDHF)~\cite{meyer1990multi, micha1994time,
fasshauer2016multiconfigurational, lode2020colloquium, lin2020mctdh}.
Although these methods are powerful tools for qualitatively capture the behavior of weakly interacting systems, they become unreliable for larger and strongly interacting systems.

With the advancement of computing power and numerical algorithms, interest has resurged in the explicit time propagation of correlated systems. Including electron correlations to predict quantum dynamics beyond mean-field approximations is a challenging task. One often resorts to a class of non-variational methods. For example, only relatively recently has the real-time time-dependent density functional theory (RT-TDDFT) approach been introduced within the local density approximation (LDA) for studying dynamic response properties~\cite{yabana1996time, isborn2007time}.
Other recent examples are time-dependent versions of multi-configuration self-consistent fields (MC-SCF) ~\cite{sato2013time, miyagi2013time, miyagi2014time, sato2015time, liu2019time}, configuration interaction (CI) \cite{krause2005time, schlegel2007electronic, sonk2011td, deprince2011emergence, luppi2012computation, lestrange2018time, ulusoy2018role}, algebraic diagrammatic construction ~\cite{cederbaum1999ultrafast, santra2002complex, feuerbacher2003complex, kuleff2007tracing, dutoi2010tracing, kuleff2014ultrafast, neville2018general}, coupled cluster (CC) theories~\cite{nascimento2016linear, sonk2011td, luppi2012computation, huber2011explicitly, pigg2012time, nascimento2017simulation, sato2018communication, kristiansen2020numerical, nascimento2019general, koulias2019relativistic, folkestad20201, skeidsvoll2020time},  dynamical mean-field theory (DMFT)~\cite{eckstein2010interaction, schiro2010time}, tensor networks~\cite{vidal2004efficient, haegeman2011time, haegeman2013post} and the density matrix renormalization group (DMRG)~\cite{cazalilla2002time}. 
Each of the aforementioned methods faces inherent limitations and/or scalability concerns. Mean-field methods such as DFT and HF encounter difficulties in describing correlation effects as explained above, while CI and CC-based techniques exhibit unfavorable scaling as the size of the system increases~\cite{hermann2022ab}.

Wave-function-based techniques offer the potential for systematically improvable accuracy and precise simulations of electronic eigenstates and are routinely used to capture ground states of electronic systems in continuous space using variational Monte Carlo (VMC). Recently, great progress has been made in the design of variational wave function ansatze, by the introduction of neural quantum states (NQS) with variational Monte Carlo~\cite{carleo2017solving}, producing impressively accurate solutions to the electronic TISE~\cite{pfau2020ab, hermann2020deep, von2022self, pescia2023message, scherbela2024towards, gao2023generalizing, romero2024spectroscopy, wu2023variational, wilson2023neural}.
NQS promise to provide high-accuracy wave functions, capturing the entire entanglement of the system, at an affordable computational cost~\cite{hermann2022ab}.
However, their application to study quantum dynamics, where correlations are expected to be essential, remains an open problem. Indeed, to the best of our knowledge, there exists no prior work that has successfully extended VMC to capture the real-time evolution of electronic systems in continuous space, despite its possible impact on quantum chemistry and condensed matter.

In this work, we go beyond the mean-field approximation and introduce a novel technique for variational time-dependent wave functions that can capture many-body correlations. We use time-dependent variational wave functions with Jastrow functions and backflow transformations to accurately reproduce the time-dependent wave functions of benchmark experiments, including the solvable harmonic-interaction model, diatomic molecules in an intense laser field, and quenched quantum dots. 

\section{Time-dependent quantum many-body wave functions}
\begin{figure*}[tbh]
    \centering
    \includegraphics[width=1.0\textwidth]{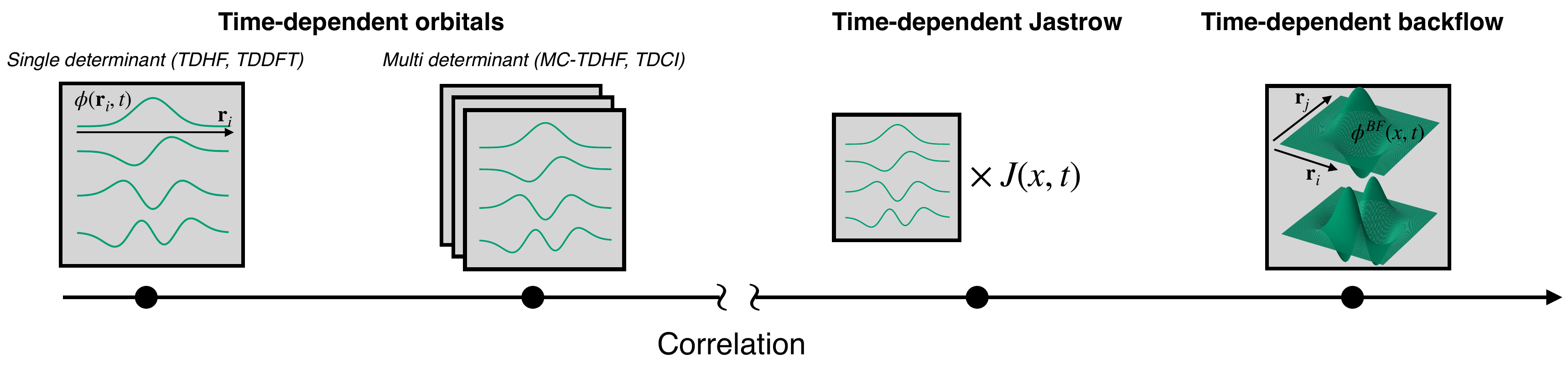}
    \caption{Schematic representation of different approaches to capture correlations. Single-determinant approaches with single-particle orbitals are mostly applicable where strong correlations can be neglected. Correlations can be built in using a polynomial number of determinants, a multiplicative time-dependent Jastrow factor that depends on all the electron positions, or in the most powerful case by using a determinant with higher-dimensional many-body time-dependent orbitals using backflow transformations.}
    \label{fig:sketch_methods}
\end{figure*}

Our aim is to capture the time evolution of $\ket{\Psi(t)}$ according to Eq.~\eqref{eq:tdse}, induced by $\hat{H}(t)$. For electron configurations $x = \left[\vec{r}_1 ,..., \vec{r}_N\right]$ in position space, the Hamiltonian reads 
\begin{align}
\hat{H}(t) = -\frac{1}{2}\sum_{i=1}^N \nabla_{\vec{r}_i}^2 + V(x, t) ,
\end{align}
with time-dependent potential $V$. We will consider spin-independent Hamiltonians and assume that all electrons have a fixed spin-projection quantum number $\sigma_i \in \{\uparrow,\downarrow\}$. 

\subsection{Variational time evolution}
We approximate the time-evolving state $\ket{\Psi(t)}$ using a parametrized Ansatz $\ket{\Phi(\theta(t))}$ with time-dependent parameters $\theta(t) = \left[\theta_1(t),..., \theta_{N_p}(t)\right]$, reducing the problem to finding the optimal $\theta(t)$ such that $\ket{\Psi(t)} \approx \ket{\Phi(\theta(t))}$. In Section~\ref{sec:tdvp}, we illustrate how the trajectories $\theta(t)$ are obtained using the time-dependent variational principle \cite{mclachlan1964time, haegeman2011time, haegeman2013post}. 
We discuss the generalization of MacLachlan's variational principle into a framework for correlated many-body wave functions, called tVMC, in Section~\ref{sec:tvmc}.
This has recently also been demonstrated for bosonic systems~\cite{carleo2017unitary, carleo2012localization}, quantum many-body spin systems arising in condensed matter physics~\cite{ haegeman2016unifying, schmitt2020quantum, gutierrez2022real, gartner2022time, sinibaldi2023unbiasing, medvidovic2024variational} and for modeling fermionic lattice Hamiltonians (the Fermi-Hubbard model)~\cite{ido2015time}. 
Our work is concerned with its application to ab initio many-electron systems in continuous space, which is relevant to quantum chemistry and condensed matter and which has remained elusive so far. \changed{Additionally, in Section \ref{sec:tretvmc}, we introduce a novel approach that combines tVMC with Taylor-root expansions to achieve precise evolution of parameterized quantum states, further enhancing the method’s capabilities in this new context, especially in conjunction with neural network ansatze.}

\subsection{Variational wave-function models}
Since the Hamiltonian operator is Hermitian and time-reversal symmetric, one can restrict oneself to real wave functions when solving the static TISE problem~\cite{neklyudov2023wasserstein}. However, to model quantum dynamics, complex wave functions are generally required.
We will introduce variational models in terms of electron positions $[\vec{r}_1, ...,\vec{r}_N]$ in a continuous position space. The wave function ansatz must obey the correct permutation symmetry under the exchange of particles, that is, it must be made antisymmetric under electronic permutations. To this end Slater determinants are routinely used, constructed from a set of $M$ complex, time-dependent, single-particle mean field orbitals $\mathcal{M}=\{\varphi_\nu(\vec{r})\}_{\nu=1}^{M}$ (we drop the parameter dependence to simplify the notation). We can go beyond this mean-field approximation and capture many-body correlations using, for example, a complex symmetric Jastrow factor dependent on time $J(x, t)$ that depends on the complete many-body configuration $x$~\cite{carleo2012localization}.
To capture additional correlations and change the nodal surface dictated by the choice of single-particle orbitals, we transform the orbitals into a set of multi-electron orbitals introducing a ``time-dependent backflow transformations'' (tBF), $ \varphi_\mu(\vec{r}_i, t) \to \varphi_\mu^{BF}(\vec{r}_i, x, t)$~\cite{ido2015time}. Hence, our variational time-dependent quantum many-body wave function reads~\footnote{Alternative approaches exist to change the nodal surface, such as the introduction of hidden fermions~\cite{robledo2022fermionic} and pair-orbitals with Pfaffians or geminals~\cite{kim2023neural, lou2023neural}.}
\begin{align}
    \Phi(x,t) = \det \left[ \varphi_\mu^{BF}(\vec{r}_i, x, t) \right] e^{J(x, t)} \label{eq:model_general}.
\end{align}
By using time-dependent backflow transformations, our model obtains high expressive power, which allows us to represent states that are significantly more correlated compared to those obtained by mean-field calculations. For a detailed description of the model, we refer to Appendix~\ref{sec:model_details}. In Fig.~\ref{fig:sketch_methods}, we provide a schematic overview of how the time-dependent Jastrow and backflow transformations compare with traditionally used time-dependent mean-field approaches.

An alternative and common choice is to forgo the direct description of the wave function in continuous space and introduce a finite basis set. In this case (second quantization), the wave function Ansatz does not need to be constrained to fulfill the correct permutation symmetry, since it is automatically captured by the fermionic creation/annihilation operators and their anti-commutation relations~\cite{szabo_modern_1996}. Instead of positions, states are represented in the occupation number basis $x=\left[n_{1}, ..., n_{M}\right]$ (with $n_i \in \{0,1\}$) for a given set of $M$ mean field orbitals $\mathcal{A} = \{\phi_\mu\}$ (for example, based on Gaussian orbitals or similar). 
We parameterize $\Psi(n_1, ..., n_M, t)$ directly with a flexible variational function, such as a neural network, with a set of time-dependent parameters $\theta(t)$, without explicit anti-symmetrization with determinants~\cite{choo2020fermionic}. In particular, we use a complex-valued, time-dependent Restricted Boltzmann Machine (RBM) as in Ref.~\cite{carleo2017solving} (see Appendix~\ref{sec:model_details} for details).

\section{Results}
\subsection{Interacting fermions in one dimension}
\begin{figure}[bt]
    \centering
 \includegraphics[width=0.5\textwidth]{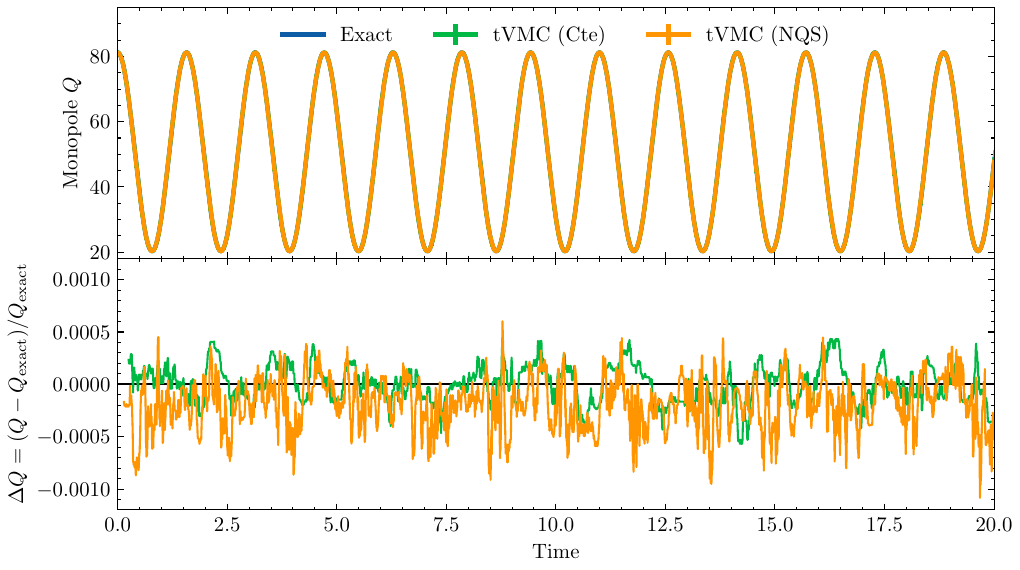}
    \caption{Monopole $Q$ for the harmonic interaction model with $30$ particles, subject to a quench of the harmonic confinement and a time-dependent interaction strength. (top panel) We show the predictions with tVMC using the ($i$) time-dependent constants Ansatz in green, and ($ii$) the neural quantum state in orange, both introduced in the main text. We compare to the exact solution in blue. The curves are overlapping and therefore hardly distinguishable. \changed{(bottom panel) The relative error $\Delta Q$ on the predicted monopole, averaged over a rolling window of $\Delta t=0.2$ to reduce the effect of statistical noise.}}
    \label{fig:him_monopole}
\end{figure}
To demonstrate the validity of our approach to capture the time-dependent state of many-body systems, we start with studying the exactly solvable harmonic interaction model in one dimension, describing harmonically confined particles interacting via a harmonic potential:
\begin{align}
    V(x, t) &=  \sum_{i=1}^N \left[ \frac{1}{2} \omega(t)^2  
    \vec{r}_i^2 + \frac{g(t)}{2}\sum_{i>j}^N (\vec{r}_i - \vec{r}_j)^2 \right] ,
\end{align}
where $\omega$ is the trap frequency and $\vec{r}_i \in \mathbb{R}$. 
In particular, we simulate the dynamics of the system subject to a trap quench, for which an analytical solution is known~\cite{zaluska2000soluble, gritsev2010scaling}. 
When the trap frequency is quenched from $\omega_0 \to \omega_\mathrm{f}$ at time $t=0$, the particles exhibit a breathing mode with period $T=\pi/\omega_\mathrm{f}$ for a well-chosen time-dependent interaction strength $g(t)$. In particular, the time-evolved state reads (up to irrelevant global factors)
\begin{align}
    \Psi(x, t) &= \det \left[\mathcal{V}\left(\frac{x}{L(t)}\right)\right] e^{-J_\textrm{well}(x, t)}e^{-J_\textrm{int}(x, t)} \label{eq:himexact_t} \\
    J_\textrm{well}(x, t) &= \alpha(t) \sum_{i=1}^N \vec{r}_i^2, \quad J_\textrm{int}(x,t) = \beta(t) \left(\sum_{i=1}^N \vec{r}_i\right)^2 
\end{align}
when $g(t) = g/L(t)^4$~\cite{gritsev2010scaling}, with $g$ the original interaction strength at $t<0$. Here, $\mathcal{V}(x)$ is the Vandermonde matrix, equivalent to a Slater determinant with orbitals $\varphi_\mu(\vec{r}) = \vec{r}^{\mu-1}$, with $\mu=1,...,N$. The time-dependent scale functions $L(t)$, $\alpha(t)$, and $\beta(t)$ can be derived analytically and are given in Appendix~\ref{sec:him}.
To study the breathing mode with a variational method, we carry out two experiments to demonstrate that tVMC can capture fermionic time-dependent correlations: ($i$) we parameterize $\alpha(t)$, $\beta(t)$, $L(t)$ with time-dependent constants, and ($ii$) we represent the correlation contribution to the Jastrow $J_\textrm{int}$ with a neural quantum state (NQS) Ansatz. For the latter, we use two general DeepSet neural network architectures (as introduced in Ref.~\cite{pescia2022neural} to model bosonic ground states), one for $\Re \left[J_\textrm{int}\right]$ and one for $\Im \left[J_\textrm{int}\right]$, with time-dependent parameters. This choice of architecture guarantees the particle-permutation invariance of the Jastrow factor.
In Figure~\ref{fig:him_monopole} we compare the evolution of the monopole $Q = \sum_i^N\expval{ \vec{r}_i^2}$ as a function of time for $30$ fermions and $\omega_0 = 1 \to \omega_\mathrm{f} = 2$ and $g=1$, using the parameterizations mentioned above. We observe that the breathing mode and correlations are both accurately reproduced with the tVMC method, thus validating our approach.

\subsection{Molecules in a laser field}
\begin{figure}[tb]
    \includegraphics[width=0.5\textwidth]{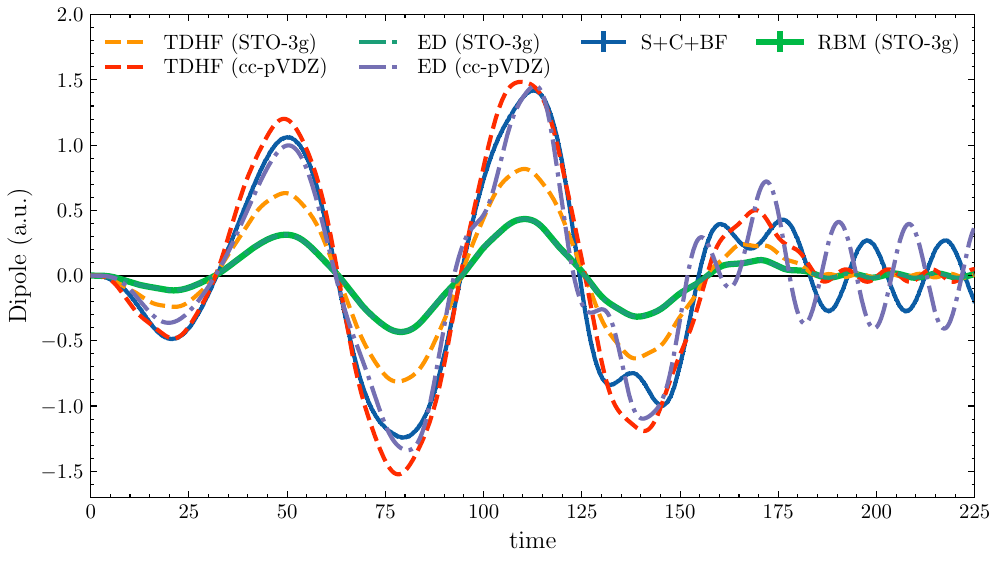}%
    \caption{Time-dependent dipole moment of $H_{2}$ in an intense, time-dependent laser field modeled with an NQS and tVMC. We show the effect of capturing correlations with a slater determinant with time-dependent neural backflow transformation and a Jastrow representing the cusp condition (S+C+BF), as well as results obtained in the STO-3g basis in second quantization (RBM). We compare to predictions from TDHF and ED.
    \label{fig:H2_tevo}}
\end{figure}
As a more complex case study, we focus on the electronic response of a diatomic molecule in an intense, time-dependent laser field. The electronic potential with $N_A$ atoms at positions $\{\vec{R}_a\}_{a=1}^{N_a}$ reads (in atomic units)
\begin{align}
    V(x, t) = \sum_{i < j}^N \frac{1}{\norm{\vec{r}_i - \vec{r}_j}} - \sum_{i}^N \sum_{a}^{N_A} \frac{Z_a}{\norm{\vec{r}_i - \vec{R}_a}} + V_{\textrm{ext}}(x, t)
\end{align}
where we introduce the charge number $Z_a$ and an external potential $V_{\textrm{ext}}(x, t)$ describing a linearly polarized and spatially homogeneous electric field (see Ref.~\cite{li2005time} and Appendix~\ref{sec:electric_profiles}). 
Hartree-Fock is known to be unable to properly describe the dissociation curve of $H_2$, especially at large distances~\cite{becca_quantum_2017}. Therefore, we will consider $H_2$ separated by twice its equilibrium distance. In Fig.~\ref{fig:H2_tevo} we show the induced dipole moment for H${}_2$ modeled in both the first and second quantization formalism.
\changed{
We include TD-HF and Exact Diagonalization (ED) results in the minimal basis set STO-3g where the number of orbitals is equal to the number of electrons. Here, $3$ Gaussian primitive functions are used to approximate each Slater-type orbital.
Since the electric field induces an increased delocalization of the orbitals, we show results for TD-HF and ED in a larger basis set cc-pVDZ, which includes correlation-consistent polarized valence basis sets and uses successively larger shells of polarization (correlating) functions.
}

The initial state of the various approaches is their respective approximation to the ground state of $H_2$, which slightly differ depending on the accuracy of the method. The ground state results are summarized in Appendix~\ref{sec:H2_gs}.
The effect of correlations (i.e.\ ED versus HF predictions) can be observed mainly in oscillation amplitudes, as well as the interference behavior superimposed on the electric-field-induced oscillatory behavior.
In contrast to TD-HF in the same basis set, tVMC with tNQS reproduces the ED results, thereby demonstrating that tNQS can accurately capture the electron correlations, even in a limited basis set. In continuous space, we use a neural network ansatz, inspired by PauliNet~\cite{hermann2020deep} and neural backflow transformations~\cite{luo2019backflow}, using a modified version of the powerful particle-attention backflow transformations recently introduced in Ref.~\cite{pescia2023message} for applications to the homogeneous electron gas. The latter has recently also found successful applications to quantum materials~\cite{luo2023pairing} nuclear matter~\cite{gnech2023distilling}, and ultra-cold dilute matter~\cite{kim2023neural}. The details of the model are given in Appendix~\ref{sec:model_first_quant_molec}. First, as a validation, we observe that tVMC with a pure mean-field Slater determinant reproduces the predictions from TD-HF in the same basis, although tVMC scales more favorably with the size of the basis set, i.e.\ $\order{M}^3$ compared to $\order{M^4}$ (see Appendix~\ref{sec:scaling} on scalability). By including a time-dependent backflow and therefore electron correlations, the time evolution of the dipole changes compared to mean field, which prominently manifests itself as a damping effect on the oscillations.

\subsection{Quenched electronic quantum dot}\label{sec:quantumdot}

\begin{figure}[tb]
    \centering
    \includegraphics[width=0.5\textwidth]{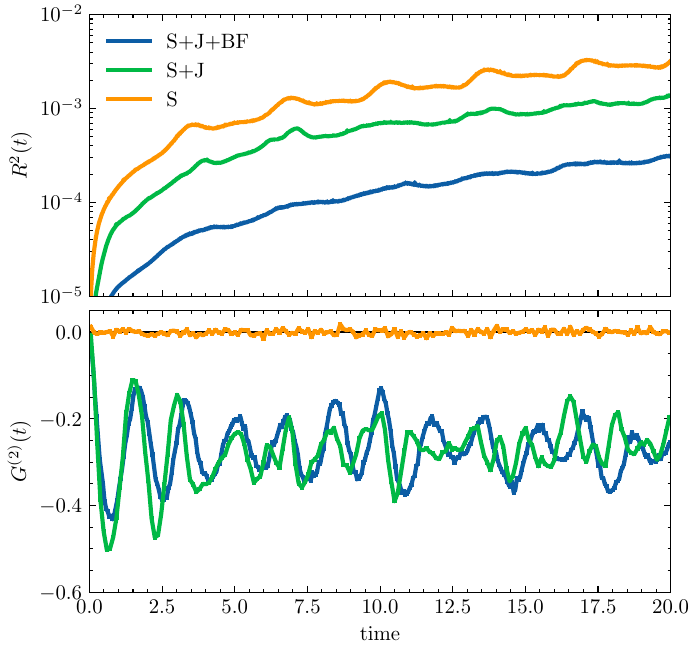}
    \caption{The integrated $R^2(t)$ error in Eq.~\eqref{eq:R2} (top panel) and pair correlation  $G^{(2)}(t)$ in Eq.~\eqref{eq:integrated_G2} (bottom panel) as a function of time for a fully polarized quantum dot with $N=6$ and subject to a quench $\kappa(t) = 1 \to 2$ at $t=0$. We compare with predictions using tVMC with a single Slater determinant (S), a Slater-Jastrow model (S+J), and a Slater-Jastrow-Backflow (S+J+BF) model.\label{fig:qg_g2_1_2}\label{fig:quantumdot_R2}}
\end{figure}

As a final demonstration of the capabilities of our method, we model the behavior of confined electrons in two dimensions, within a harmonic well. This system, also known as a quantum dot, describes electrons propagating in a semi-conducting material. It exhibits intensified Coulomb interactions due to a strong dielectric confinement~\cite{lhuillier2015two, diroll2020colloidal, richter2017nanoplatelets}, which influence crucial optoelectronic properties~\cite{planelles2021simple, jacak2013quantum, bimberg1999quantum, chakraborty1999quantum}. The simulation of these Coulomb-induced correlations is vital for describing these properties~\cite{mcdonald2013theory, serra2003breathing, bauch2010quantum, shumway2001correlation} and the system therefore serves as an ideal test-bed to compare to less expressive mean-field methods.
 
The interaction and confining potential in this system reads
\begin{align}
    V(x, t) &=  \sum_{i=1}^N  \frac{1}{2} \omega^2  
    \vec{r}_i^2 + \sum_{i < j}^N \frac{\kappa(t)}{\norm{\vec{r}_i - \vec{r}_j}} ,
\end{align}
where $\omega$ is the trap frequency, and $\kappa > 0$ is the interaction strength~\cite{bauch2010quantum}, related to the dielectric constant of the semiconductor~\cite{xie2021ab}. We consider the evolution of the electronic ground state of the Hamiltonian with $\kappa(t<0)=1$ after a quench \changed{that induces a sudden screening of the mutual Coulomb interaction due to a change in the static dielectric constant of the semiconductor. The latter is modeled} by an abrupt doubling of the effective interaction strength $\kappa(t\geq 0) = 2$ at $t=0$, while fixing the confining potential $\omega = 1$. 

Although one can again observe breathing modes in the electric monopole (see Fig.~\ref{fig:quantumdot_monopole_1_2} in Appendix~\ref{sec:quantumdot_monopole}), we focus here on observables that specifically highlight the correlations in the system.
To highlight the effect of correlations in this system, we include the connected pair-correlation function $G^{(2)}$ in Fig.~\ref{fig:qg_g2_1_2}, which isolates correlation effects and is identically zero in the mean-field limit (see Section~\ref{sec:pair_correlation} for the definition). We observe that, even though correlations do not have a quantitative effect in the ground state, they significantly contribute shortly after the quench and remain present throughout the time evolution. 
In Fig.~\ref{fig:quantumdot_R2} we also show the integrated error measure $R^2$, introduced in Section~\ref{sec:tdvp_extra}, to quantify the degree to which the models satisfy the TDSE. We conclude that a more expressive wave function also yields significantly lower integration errors and \changed{is therefore closer to the exact solution of the time-dependent Schrödinger equation}. The presence of correlations explains the rapid increase in integration error for the mean-field model in Fig.~\ref{fig:quantumdot_R2}, as the latter does not capture the strong correlations induced by the change in interaction strength. Furthermore, allowing the nodal surface to vary over time (using backflow transformations) yields more accurate and truthful dynamics.

\begin{figure}[tbh]
    \centering
    \includegraphics[width=1.0\linewidth]{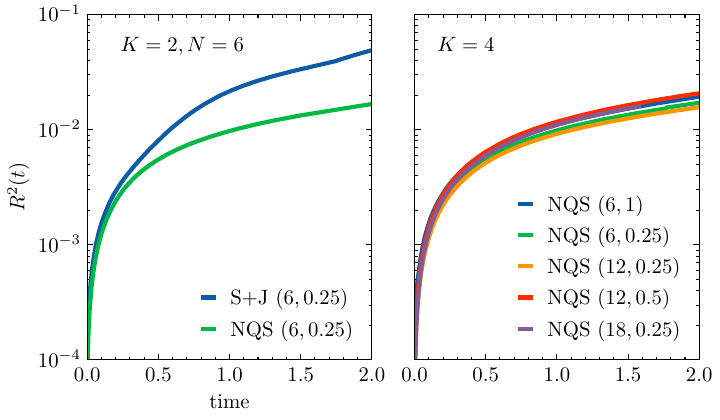}
    \caption{The integrated $R^2$ for the same quantum dot quenched experiment as in Fig.~\ref{fig:quantumdot_monopole_1_2}, using the tre-tVMC approach with given order $K$. Between brackets, we indicate the number of electrons and the time step: $(N, \dt\times 10^2)$. We compare the results using the basis expanded Slater-Jastrow model (S+J) with $K=2$ TRE to a fully neural-network-based wave function ansatz (NQS). }
    \label{fig:quantumdot_R2_implicit}
\end{figure}

\begin{figure}[tbh]
    \centering
    \includegraphics[width=0.9\linewidth]{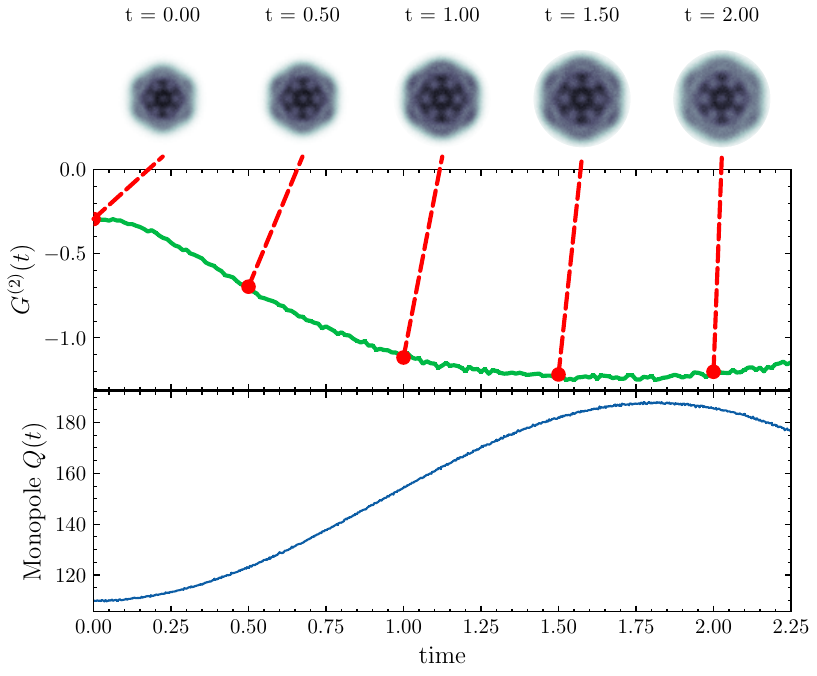}
    \caption{Integrated pair correlation $G^{(2)}(t)$ and monopole $Q(t)$ as a function of time. We show the results obtained with the NQS wave function for the 2d quantum dot with $N=18$ electrons (see Fig.~\ref{fig:quantumdot_R2_implicit}). In the top row, we show snapshots of the single-body density $\rho(\vec{r})$ .}
    \label{fig:quantumdot_timeline}
\end{figure}

\changed{We also demonstrate the use of neural quantum states to capture the correlated electronic time-dependent wave function of the quantum dot quench. To this end, we use a novel higher-order method, we call tre-tVMC, detailed in Section~\ref{sec:tretvmc}. In Figure \ref{fig:quantumdot_R2_implicit}, we compare the integration error for an NQS ansatz with the basis-expanded wave function for $N=6$ (see Appendix~\ref{sec:model_details} for model details). We observe that the NQS-based ansatz is more expressive due to the flexible neural orbitals, and better captures the electron correlations through the neural Jastrow factor. We also show the results for larger systems up to $N=18$ electrons, using the higher-order Taylor-root expansion scheme with a local error $\mathcal{O}(\dt^5)$ (denoted $K=4$). This demonstrates that our approach yields accurate results for various system sizes. The largest many-electron system of $N=18$ fermions is shown in more detail in Fig.~\ref{fig:quantumdot_timeline} where we observe strong variations in the single-body density over time. Hence, during the quench the delocalization of the orbitals oscillates, thereby suggesting that wave function models without basis dependence are most appropriate to capture the time-dependent wave function of the electron systems.
}

\section{Discussion\label{sec:conclusions}}
We have introduced a variational approach to capture the dynamics of quantum many-electron systems, specifically addressing the challenges posed by real-time ab initio electron dynamics. Our methodology involves the use of correlated time-dependent variational wave functions, which surpass the typically adopted mean-field approximations. In particular, we adopted here the time-dependent Jastrow function and the backflow transformation to capture electronic correlations. 
We discussed how to time-evolve the variational wave functions using the time-dependent variational Monte Carlo (tVMC).

We demonstrated the effectiveness of our approach through applications to three distinct systems: the solvable harmonic interaction model, diatomic molecules in intense laser fields, and a quenched quantum dot. In the harmonic interaction model, we successfully reproduced the breathing mode induced by a trap quench, underscoring the method's accuracy in capturing correlated many-body system dynamics. Transitioning to diatomic molecules in intense laser fields, our approach demonstrated versatility in both first and second quantization. In both cases, we demonstrated that tVMC can be used to time-evolve expressive models such as Neural Quantum States, as well as the more traditional basis-expanded models. \changed{We also introduce a novel alternative to tVMC, called tre-tVMC, that is particularly powerful when using expressive neural-network wave functions in continuous space.}
In first quantization, the incorporation of neural backflow transformations showcased a promising avenue for simulating complex molecular systems in external fields. Finally, our approach excelled in predicting observables for the quenched quantum dot, emphasizing the significance of capturing many-body correlations in the dynamics of strongly interacting electronic systems.
In conclusion, our variational time-dependent wave function approach, as demonstrated across different systems, presents a promising direction for advancing real-time electron dynamics, offering a balance between accuracy and computational efficiency in capturing the intricate dynamics of quantum many-electron systems.

The presented work opens avenues for further exploration and improvement. One avenue is the extension of our approach to larger and more complex systems. Additionally, exploring more powerful ansatzes may provide further improvements in accuracy and efficiency. Furthermore, the applicability of our method to more diverse quantum many-electron systems, including materials and chemical reactions, could uncover new insights into their dynamic behaviors.

\section{Methods}\label{sec:methods}

\subsection{Variational time evolution with TDVP}\label{sec:tdvp}
The time evolution of a state $\ket{\Phi(\theta(t))}$ over a small time interval $\dt$ described by the TDSE can be approximated as $\ket{\tilde{\Psi}(t + \dt)} = \left(1-i \dt\hat{H}\right) \ket{\Phi(\theta(t))}$. A variational approach projects $\ket{\tilde{\Psi}(t + \dt)}$ back onto the trial state manifold or requires finding the $\theta(t+\delta t)$ that optimally approximates the state evolved in time. These approaches are known as ``time-dependent variational principles'' (TDVP).
Various methods are available for this projection and include the Dirac-Frenkel (DF) and MacLachlan (McL) variational principles. In particular, McL reduces to DF for holomorphic wave function parameterizations. However, in general, DF conserves the energy of the system subject to time-independent Hamiltonians, in contrast to the non-symplectic McL approach. For a discussion and comparison of the various TDVPs available, we refer to Refs.~\cite{broeckhove1988equivalence, hackl2020geometry, yuan2019theory}. Furthermore, explicit projection methods have recently been introduced~\cite{sinibaldi2023unbiasing, gutierrez2022real, gacon2023variational}. For simplicity, we will assume in this section that all parameters are complex and the wave function Ansatz is holomorphic.
We minimize the (Fubini-Study) distance between the time-evolved state $\ket{\tilde{\Psi}(t + \dt)}$ and the parametrized state $\ket{\Phi(\theta(t+\dt))}$, setting $\theta(t+\dt) = \theta(t)+\dot{\theta}(t)\dt$ for small time-steps $\dt$, to obtain a first-order differential equation for the variational parameter velocities $\dot{\theta}$,
\begin{align}
    \sum_{k'}^{N_p} G_{k, k'}(t) \dot{\theta}_{k'}(t) = -i F_k(t) \label{eq:tdvp},
\end{align}
with the Quantum Geometric Tensor $G$ (QGT) and the energy gradients $F$ given by
\begin{align}
    G_{k, k'}(t) &= \braket{\pder{}{\theta_k}\Phi(\theta(t))}{\pder{}{\theta_{k'}}\Phi(\theta(t))} ,\label{eq:qgt}\\
    F_{k}(t) &= \mel{\pder{}{\theta_k}\Phi(\theta(t))}{\hat{H}(t)}{\Phi(\theta(t))}.\label{eq:forces}
\end{align}

\subsection{Time-Dependent Variational Monte Carlo (tVMC)}\label{sec:tvmc}
Although MacLachlan's and Dirac-Frenkel's variational principle were originally introduced in conjunction with mean-field states, one can generalize the approach to incorporate many-body correlations. Therefore, we formulate the variational time evolution using Monte Carlo estimates to integrate the system over the exponentially large Hilbert space, as we will demonstrate in this section.
We aim to capture the dynamics of the probability amplitudes $\Psi(x, t) \equiv \braket{x}{\Psi(t)} \in \mathbb{C}$, which we parameterize as $\Phi(x, \theta(t)) \equiv \braket{x}{\Phi(\theta(t))}$. Here, $x$ represents a set of continuous electron positions and spins, or an occupation configuration in a given basis set. Since the Hilbert space scales exponentially with the system size, we resort to Monte Carlo estimates of the quantities in Eqs.~\eqref{eq:qgt} and \eqref{eq:forces}. The time-dependent variational principle in combination with Monte Carlo is referred to as time-dependent Variational Monte Carlo (tVMC)~\cite{carleo2012localization,carleo_light-cone_2014,becca_quantum_2017}. 
For a given trial wave function, the energy can be computed using (we drop the time dependence to simplify the notation),
\begin{align}
    E &= \mathbb{E}_{x \sim \abs{\Phi(x,\theta)}^2}\left[ \Eloc(x) \right] , \label{eq:tvmc_energy}
\end{align}
where we introduced the local energy $\Eloc(x) = \tfrac{\left[\hat{H}\Phi\right](x, \theta)}{\Phi(x, \theta)}$. 
Furthermore, by introducing the log derivative of the wave function with respect to parameter $\theta_k$: $O_k(x) = \partial_{\theta_k} \log \Phi(x, \theta)$, we obtain the following estimators
\begin{align}
    G_{k, k'} &= \mathbb{E}_{x \sim \abs{\Phi (x,\theta)}^2} \left[O_k^*(x) \Delta O_{k'}(x)\right] \label{eq:tvmc_qgt}, \\
    F_k &= \mathbb{E}_{x \sim \abs{\Phi (x,\theta)}^2} \left[ O_k^*(x) \Delta \Eloc(x)\right] \label{eq:tvmc_force},
\end{align}
where $\Delta \Eloc(x) =  \Eloc(x) - E$ and $\Delta O_k(x) = O_k(x) - \mathbb{E}_{x' \sim \abs{\Phi (x',\theta)}^2} \left[O_k(x')\right]$. We refer to Appendix~\ref{sec:details_computation_hamiltonian_state} for more details.

\changed{
After estimating the energy derivatives and quantum geometric tensor using Eqs.~\eqref{eq:tvmc_force} and ~\eqref{eq:tvmc_qgt}, we solve the linear system of equations in Eq.~\eqref{eq:tdvp} using the regularized singular value decomposition introduced in Refs.~\cite{medvidovic2024variational, medvidovic2024neural}. An unbiased estimate of the QGT and a stable and accurate inversion of the latter is essential to time-evolve the parametrized state with tVMC, and can become challenging in the large-parameter limit relevant for highly expressive neural-network-based wave functions. For this reason, in the next section we introduce a complementary approach more suitable when considering such regime. 
}

\subsection{Taylor-root expansion tVMC (tre-tVMC)}\label{sec:tretvmc}
We introduce an additional method to evolve a parametrized quantum state using variational Monte Carlo.
Our central goal is again to maximize the fidelity between the true time-evolved state $\ket{\Psi(t+\dt)}$, and a state on the variational manifold $\ket{\Phi(\theta)}$.
For this task, we first consider a Taylor expansion consistent to order $K$ in $\dt$.
Rather than evaluating higher orders in $\hat{H}$ directly (which is computationally expensive, especially in continuous space), we compute these through sequential optimizations. One possible approach is to variationally compress the states in the Krylov basis $\{1, \hat{H}, \hat{H}^2,...\} \ket{\Psi(t)}$, yet this proves challenging in practice, since these basis states are significantly different from the initial reference state $\ket{\Psi(t)}$. A more convenient method is to target intermediate states that are at most $\order{\dt}$ different from $\ket{\Psi(t)}$, thereby greatly reducing the computational burden.
With this idea in mind, we factorize the time evolution operator as
\begin{align}
    e^{-i \dt \hat{H}} &= \prod_{k=1}^K \hat{R}_k + \order{\dt^{K+1}}, \label{eq:prod}
\end{align}
and set $\hat{R}_k = 1 - i c_k \dt \hat{H}$ with constants $c_k \in \mathbb{C}$ to be determined. The latter can be obtained by equating Eq.~ \eqref{eq:prod} to the Taylor expansion of the propagator. In other words, we rewrite the propagator in terms of the roots of the Taylor polynomials associated with the exponential function (which we refer to as the \emph{Taylor roots expansion})~\cite{conrey1988zeros}.
For example, for $K=2$ this results in $c_k = (1\pm i)/2$. Hence, we sequentially optimize the fidelity for the evolution operators
\begin{align}
    \hat{R}_1 &= 1 - i \dt \left(\tfrac{1+i}{2}\right) \hat{H} , \qquad\hat{R}_2 = 1 - i \dt \left(\tfrac{1-i}{2}\right) \hat{H} .
\end{align}
In practice, we obtain a first set of optimal parameters $\theta_1^*$ by optimizing $\mathcal{F}\left(\smash{\hat{R}_1\ket{\Psi(t)},\ket{\Phi(\theta_1)}}\right)$, and sequentially obtain a second set of parameters $\theta_2^*$ by optimizing $\mathcal{F}\left(\smash{\hat{R}_2\ket{\Phi(\theta_1^*)},\ket{\Phi(\theta_2)}}\right)$, resulting in a state $\ket{\Phi(\theta_2)}$ that approximates $\ket{\Psi(t+\dt)}$ consistently to second order in $\dt$: $\ket{\Psi(t)} \overset{\hat{R}_1}{\longrightarrow} \ket{\Phi(\theta_1^*)} \overset{\hat{R}_2}{\longrightarrow} \ket{\Phi(\theta_2^*)} \approx \ket{\Psi(t+\dt)} $.
The latter yields an integration consistency similar to the Heun integration scheme. Higher-order integrators can easily be obtained in a similar way by solving Eq.~\eqref{eq:prod} for higher $K$ (see Appendix~\ref{sec:tretvmc_discussion}). We optimize the fidelity using the MC estimator discussed in Appendix~\ref{sec:tretvmc_discussion}, where we also introduce a set of useful tricks to accelerate the evolution and optimization.

With this approach and for a fixed time step $\dt$, we can now simulate the dynamics of a quantum state consistently to order $\dt^K$ at $K$ times the computational cost as a first-order approach (corresponding to $K=1$ with $c_1 = 1$). We dub this method ``tre-tVMC'', for Taylor roots expansion tVMC. Our approach improves the stability of the time evolution by avoiding using the TDVP principle, which requires an accurate inversion of the QGT. Therefore, the method is more suitable in the large-parameter regime where it can be turned into a method that scales linearly with the number of parameters involved~\cite{chen2024empowering}.
On the other hand, every time step requires a set of optimizations that can render tVMC significantly more efficient for smaller models.

Other options than the Taylor expansion are, in principle, possible. For example, Ref.~\cite{sinibaldi2023unbiasing} considered Trotter expansions which yield a set of unitary operators, thereby simplifying the fidelity estimator for each Trotter factor. However, in MC the operators in the Trotter expansions must be estimated \emph{sequentially} for each time step, which induces a scaling with the number of particles involved and therefore limits the time evolution to smaller system sizes. Furthermore, Trotterization is significantly more involved for ab initio continuous-space Hamiltonians with unbounded potentials, such as the Coulomb potential~\cite{yan2015incorporating}. Additionally, by evolving with operators containing the full Hamiltonian $\hat{H}$, we can also maintain the symmetry of the Hamiltonian in time. 

\subsection{Integrated infidelity}\label{sec:tdvp_extra}
We can estimate the error induced by the restricted capability of the variational model to represent $\ket{\Psi(t+\delta t)}$, by introducing the residuals~\cite{carleo2017solving, schmitt2020quantum} 
\begin{align}
    r^2(t) &= \mathcal{D}^2\left(\ket{\Psi(t+\delta t)}, \ket{\Phi(\theta(t+\delta t))}\right) / (N \dt)^2 \\
    &\approx N^{-2} \left[\variance{\hat{H}}{\Phi(\theta(t+\delta t))} + \dot{\theta}^\dagger G \dot{\theta} +2 \Im \left(F^\dagger \dot{\theta} \right)\right],
\end{align}
where $\mathcal{D}(\ket{\Psi}, \ket{\Phi})$ represents the Fubini-Study distance between quantum states $\ket{\Psi}$ and $\ket{\Phi}$, and the last line is obtained through a second-order consistent expansion in $\dot{\theta}$ and 
$\variance{\smash{\hat{H}}}{\Psi} = \mathbb{E}_{x \sim \abs{\Psi}^2}\left[E_{\textrm{loc}}^*(x) \Delta E_{\textrm{loc}}(x) \right]$. Furthermore, we introduce the integrated fidelity
\begin{align}
    R^2(t) = \int_0^t \dd t' r^2(t'). \label{eq:R2}
\end{align}

\subsection{Pair correlation function}\label{sec:pair_correlation}
To identify beyond-mean-field correlations, we introduce a pair-correlation function that vanishes for pure mean-field states. We introduce the pair correlation function for a general state $\ket{\Psi}$ (with position vectors $\vec{x} \neq \vec{y} \in \mathbb{R}^{d}$, and assuming a fully polarized system such that we can ignore spin)
\begin{align}
    g^{(2)}(\vec{x}, \vec{y}) = \phantom{+}&\mel{\Psi}{\phi^\dagger (\vec{x}) \phi (\vec{x}) \phi^\dagger (\vec{y}) \phi (\vec{y})}{\Psi} / \mathcal{N} \nonumber \\
    - &\mel{\Psi}{\phi^\dagger (\vec{x}) \phi (\vec{x})}{\Psi} \mel{\Psi}{\phi^\dagger (\vec{y}) \phi (\vec{y})}{\Psi}  / \mathcal{N}^2 \nonumber \\
    + &\mel{\Psi}{\phi^\dagger (\vec{x}) \phi (\vec{y})}{\Psi} \mel{\Psi}{\phi^\dagger (\vec{y}) \phi (\vec{x})}{\Psi}  / \mathcal{N}^2 .
\end{align}
where $\mathcal{N} = \braket{\Psi}$ is the normalization, and where we introduced the field creation and annihilation operators $\{\phi^\dagger(\vec{x}), \phi(\vec{x})\}$ at position $\vec{x}$.
We define the fully integrated pair correlations function as
\begin{align}
    G^{(2)} = &\int \dd \vec{x} \dd \vec{y}  g^{(2)}(\vec{x}, \vec{y}). \label{eq:integrated_G2}
\end{align}
We obtain the Monte Carlo estimator (see Appendix~\ref{sec:pair_correlation_derivation} for a detailed derivation)
\begin{align}
    G^{(2)} &= N^2   \mathbb{E}_{ \substack{x\sim \abs{\Psi}^2  \\ x'\sim \abs{\Psi}^2}} \left[ \frac{\Psi(\vec{r}_1', \vec{r}_2,...,\vec{r}_N)}{\Psi(\vec{r}_1, \vec{r}_2,...,\vec{r}_N)} \frac{\Psi(\vec{r}_1, \vec{r}_2',...,\vec{r}_N')}{\Psi(\vec{r}_1', \vec{r}_2',...,\vec{r}_N')} - \frac{1}{N}\right]  .
\end{align}
For a mean-field state, the pair-correlation function vanishes $G^{(2)} \equiv 0$.

\section{Data Availability}
\changed{The data can be generated using the publicly available code. The models and observables can be found on the project's GitHub page \url{https://github.com/cqsl/electron-tvmc}.}

\section{Code availability}
\changed{The simulations in this work were carried out using NetKet~\cite{vicentini2022netket, carleo2019netket}, which relies on Jax~\cite{jax2018github}, and MPI4Jax~\cite{hafner2021mpi4jax}. For the molecular system, we used PySCF~\cite{sun2018pyscf} for benchmarking purposes and to construct the Hamiltonian in second quantization. 
The code implementing the tre-tVMC approach is publicly available \url{https://github.com/cqsl/tre-tVMC} and Ref.~\cite{jannes_nys_2024_13273354}.}

\bibliography{biblio.bib}

\begin{acknowledgments}
    The authors would like to thank Markus Holzmann  for useful and inspiring discussions. 
    This work was supported by Microsoft Research, and by the Swiss National Science Foundation under Grant No. 200021\_200336, by the NCCR MARVEL, a National Centre of Competence in Research, funded by the Swiss National Science Foundation (grant number 205602). A.S.\ is supported by SEFRI under Grant No.\ MB22.00051 (NEQS - Neural Quantum Simulation).
\end{acknowledgments}

\newpage
\onecolumngrid
\appendix
\newpage

\section{Hamiltonian and Hilbert space in second quantization\label{sec:details_computation_hamiltonian_state}}

The Hamiltonian operator in second quantization reads
\begin{align}
    \hat{H}(t) &= \sum\limits_{\substack{i,j \\ \sigma}} h_{ij}(t) \hat{a}_{i,\sigma}^\dagger \hat{a}_{j,\sigma} + \sum\limits_{\substack{i,j,k,l \\ \sigma, \sigma'}} h_{ijkl}(t) \hat{a}_{i,\sigma}^\dagger \hat{a}_{j,\sigma'}^\dagger \hat{a}_{k,\sigma'} \hat{a}_{l,\sigma}\label{eq:hamiltonian_2nd}
\end{align}
where we have defined the one- and two-body integrals $(h_{ij}, h_{ijkl})$, and the electronic creation and annihilation $(a_{i,\sigma}^\dagger, a_{i,\sigma})$ operators corresponding to atomic orbital $i \in \{1, ..., M\}$ and spin $\sigma \in \{\uparrow, \downarrow\}$. The latter respect the anti-commutation relations $\{a_{i, \sigma}^\dagger, a_{j, \sigma'}\} = \delta_{ij} \delta_{\sigma \sigma'}$. 
The quantum state is represented in the basis of electron-mode occupation numbers of a chosen basis, 
\begin{align}
\ket{\Psi(t)} = \sum\limits_{\substack{n_{i, \sigma} \in \{0, 1\}\\ \sum_i n_{i, \sigma} = N_\sigma}} \Psi(n_{1, \uparrow}, ..., n_{M,\uparrow}, n_{1, \downarrow}, ..., n_{M,\downarrow}, t) \ket{n_{1, \uparrow}, ..., n_{M,\uparrow}, n_{1, \downarrow}, ..., n_{M,\downarrow}}
\end{align}
where $N_\sigma$ represents the number of electrons with spin $\sigma$.
We will use the shorthand notation $\ket{n}=\ket{n_{1, \uparrow}, ..., n_{M,\uparrow}, n_{1, \downarrow}, ..., n_{M,\downarrow}}$.  In practical implementations, the electronic operators and occupation numbers are mapped onto spin-degrees of freedom~\cite{jordan1993paulische, nys2022variational, vicentini2022netket}. A key ingredient to the simulation of the dynamics is the choice of orbital basis in which we express the Hamiltonian. In Appendix~\ref{sec:orbital_rotation} we introduce a basis rotation that yields highly accurate time-evolution trajectories.

\section{Variational wave function models\label{sec:model_details}}
\subsection{Wave function models in first quantization for molecules}\label{sec:model_first_quant_molec}
In first quantization, any valid variational wave function must respect the electron-permutation anti-symmetry. 
Consider a set of $N_A$ atom positions $\{\vec{R}_a\}_{a=1}^{N_A}$.
Starting from an atomic orbital basis set $\mathcal{A} = \{\varphi_\mu(\vec{r})\}_{\mu}$ we form a set of single-particle molecular orbitals $\mathcal{M}=\{\psi_\nu(\vec{r})\}_{\nu=1}^{M}$ through a time-dependent linear combination: $\psi_\mu(\vec{r}, t) = \sum_\mu c_{\mu,\nu}(t) \varphi_\nu(\vec{r})$, where $c_{\mu,\nu}(t) \in \mathbb{C}$.  We introduce a time-dependent backflow transformation based on an adapted version of our recently introduced message-passing neural backflow (MP-NQS), which is composed of the novel and powerful ``Particle Attention'' mechanism, designed to efficiently capture particle correlations~\cite{pescia2023message}. Hereby, we consider the electron-electron interaction graph and introduce the node and edge variables
\begin{align}
    \vec{x}_i^{(0)} &= \left[\norm{\vec{R}_{ia}}, \vec{R}_{ia}\right]_{a=1}^{N_A} \\
    \vec{x}_{ij}^{(0)} &= \left[\norm{\vec{r}_{ij}}, \vec{r}_{ij}, \sigma_{ij}\right]
\end{align}
where $\vec{R}_{ia} = \vec{r}_i - \vec{R}_a$ are the electron-nucleus distance vectors,  $\vec{r}_{ij} = \vec{r}_i - \vec{r}_j$ the electron-electron distance vectors, and $\sigma_{ij} = +1 (-1)$ for equal (opposite) spin pairs, and the square brackets denote concatenation. 
In the first step, we reduce the dimensions of the above-mentioned vectors through a projection
\begin{align}
    \vec{x}_i &= W_1 \cdot \vec{x}_i^{(0)} \\
    \vec{x}_{ij} &= W_2 \cdot \vec{x}_i^{(0)}
\end{align}
If the dimension of the original vectors is $\mathbf{x}_{i}^{(0)} \in \mathbb{R}^{D_1}$ and $\mathbf{x}_{ij}^{(0)} \in \mathbb{R}^{D_2}$ then we take $W_{k=1,2} \in \mathbb{R}^{N_h \times D_k}$ with $N_h$ a (smaller) chosen hidden dimension.

We transform the edge-features to query/key matrices $\mathbf{Q}_{ij} = W_{Q} \cdot \mathbf{x}_{ij}$ and $\mathbf{K}_{ij} = W_{K} \cdot \mathbf{x}_{ij}$, respectively, using weight matrices $W_{Q}, W_{K} \in \mathbb{R}^{N_h \times N_h}$ with hidden dimension $N_h$.
The key/query vectors $\vec{Q}_{ij}, \vec{K}_{ij} \in \mathbb{R}^{N_h}$ are now used to efficiently capture electron-electron correlations. To this end, we compute their dot product
\begin{align}
\boldsymbol{\omega}_{ij}&=\vec{g}\left(\sum_l \mathbf{Q}_{il} \mathbf{K}_{lj}\right).
\end{align}
(where $\vec{g}$ is an MLP with output dimension $N_h$) that filter out the relevant information from the value function $\vec{\phi}$ (an MLP) to generate a set of messages
\begin{align}
    \mathbf{m}_{ij} &= \boldsymbol{\omega}_{ij}(\mathbf{x}_{ij}) \odot \boldsymbol{\phi}(\mathbf{x}_{ij}, \vec{x}_i, \vec{x}_j)
\end{align}
Since the ``Particle Attention'' mechanism is highly efficient in capturing electron correlations, we will restrict to a single iteration of the above message mechanism and create the backflow distortion vectors
\begin{align}
    \delta\mathbf{x}_i &= \boldsymbol{f}\left( \mean_{j\neq i}\mathbf{m}_{ij} \right)
\end{align}
where $\mean$ indicates the mean.
All parameters in the above-mentioned transformation are taken to be time-dependent. The final backflow-transformed quasi-particle positions are then obtained by projecting the distortion vector back to $D=3$ dimensions
\begin{align}
    \vec{r}_i \to \vec{y}_i(\vec{X}, t) = \vec{r}_i + \vec{e}_i(t) + W_P \cdot \delta\mathbf{x}_i(\vec{X}, t)
\end{align}
where $W^P \in \mathbb{R}^{D \times N_h}$ and $\vec{e}_i(t) \in \mathbb{R}^D$ is a constant time-dependent parameter vector.

The atomic orbitals themselves consist of primitive functions, i.e.\ linear combinations of Gaussian functions with polynomial pre-factors determined by the angular momentum. 
To obtain the ground state, we vary the coefficients of the atomic orbitals in the Gaussian basis, as well as the scale of the Gaussian envelopes (we choose the 6-311g basis set, as in Ref.~\cite{hermann2020deep}). Their initialized values are obtained from Hartree-Fock, using \texttt{PySCF}~\cite{sun2018pyscf}.
Hereby, we clip the Gaussian scales ($\sim e^{-\zeta \norm{\vec{R}}_{ia}^2}$) to $\zeta\leq 5$, thereby eliminating the cusp-like behavior in the atomic orbitals. These adjustments stabilize the ground state optimization in conjunction with the added cusp condition detailed below. During the real-time evolution, we keep these parameter sets fixed.

We pay attention to satisfying Kato's cusp condition for both electron-electron and electron-nucleus interactions through an appropriate Jastrow factor $J(x)$, which is a function of the original coordinates $x$.
The following Jastrow factors are included to capture the correct cusp conditions:
\begin{align}
    J(\vec{X},t) &= \sum_i \left[\sum_{j< i}\gamma^{ee}_{ij}(\norm{\vec{r}_{ij}}, t) + \sum_{a} \gamma^{ea}_{ia}(\norm{\vec{R}_{ia}}, t)\right] \label{eq:ee_cusp}
\end{align}
where 
\begin{align}
    \gamma^{ee}_{ij}(\norm{\vec{r}_{ij}}, t)  &= -\frac{c_{ij}}{\alpha^{ee}_{ij}(t) \left[1+\alpha^{ee}_{ij}(t) \norm{\vec{r}_{ij}}\right]} 
\end{align}
with $c_{ij} = 1/2$ if $\sigma_i \neq \sigma_j$ and $c_{ij} = 1/4$ if $\sigma_i = \sigma_j$, and $\alpha^{ee}_{ij}$ a variational parameter that are different for same and opposite spin configurations~\cite{foulkes2001quantum}. Similarly, we add the electron-nucleus cusp for a nucleus with charge $Z_a$
\begin{align}
    \gamma^{ea}_{ia}(\norm{\vec{R}_{ia}}, t)  &= \frac{Z_a}{\alpha^{ea}(t) \left[1+\alpha^{ea}(t) \norm{\vec{R}_{ia}}\right]}
\end{align}

\subsection{Variational wave function in first quantization for the quantum dot}\label{sec:model_first_quant_qd}
For the quantum dot, we express the one-electron orbitals in terms of associated Laguerre polynomials (eigenstates of the 2D quantum harmonic oscillator (QHO)).
\begin{align}
    \varphi_{n,m}(\vec{r}) &= \sqrt{\frac{2n!}{2\pi(\abs{m}+n)!}} e^{im\phi} r^{\abs{m}} e^{-r^2/2} L_n^{\abs{m}}(r^2) \label{eq:laguerre_gauss_basis}
\end{align}
where $(r, \phi)$ are the polar coordinates of particle vector $\vec{r}$ in two spatial dimensions.
The time-dependent mean-field orbitals read
\begin{align}
    \psi_\mu(\vec{r}, t) = \sum_{n,m} c_{\mu,(n,m)}(t) \varphi_{n,m}(r, \phi)\label{eq:qho_sum}
\end{align}
In practice, we introduce an energy cutoff $E_c$ to the above sum Eq.~\eqref{eq:qho_sum}, based on the orbital mean-field energies $e_{n, m} = 1+\abs{m} + 2n$.

As in Ref.~\cite{zen2015ab}, we decompose the Jastrow factor in this fixed basis set $\mathcal{A} = \{\varphi_{n,m}\}$:
\begin{align}
    J(x) = \sum\limits_{\substack{i<j \\ n, m, n',m'}} a_{(n, m), (n',m')}(t) \varphi_{n,m}(\vec{r}_i)\varphi_{n',m'}(\vec{r}_j)
\end{align}
Similarly, we decompose the orbital backflow transformation $\psi_\mu(\vec{r}_i, t) \to \psi_\mu^{BF}(\vec{r}_i, x, t)$ in this basis set and obtain
\begin{align}
    \psi_\mu^{BF}(\vec{r}_i, x, t) &= \psi_\mu(\vec{r}_i, t)  \sum\limits_{\substack{j \neq i \\ n, m, n',m'}} b_{(n, m), (n',m')}(t) \varphi_{n,m}(\vec{r}_i)\varphi_{n',m'}(\vec{r}_j)
\end{align}
In the above, ${a, b, c}$ are time-dependent complex variational parameters of the model. This model is holomorphic w.r.t.\ the variational parameters. Therefore, tVMC conserves the energy of the system for constant quenched Hamiltonians~\cite{hackl2020geometry}.

With the tre-tVMC method, we also use a fully neural-orbital basis with Gaussian envelopes
\begin{align}
    \phi_\mu(\vec{r}, t) &= f_\mu(\vec{r}, t) e^{-\alpha_\mu(t) \abs{\vec{r}}^2}
\end{align}
where $f_\mu$ complex holomorphic MLPs (with GELU activation functions) and $\alpha_\mu \in \mathbb{C}$ variational parameters.
For the neural Jastrow, we use the DeepSet model that was used to model bosonic degrees of freedom in Ref.~\cite{pescia2022neural}, with an additional position input
\begin{align}
    J(x, t) = \rho\left(\sum_i^N \phi_1(\vec{r}_i, t) + \sum_{j<i}^N \phi_2(\abs{\vec{r}_{ij}}^2,t) , t\right)
\end{align}
where $\rho$, $\phi_1$ and $\phi_2$ are holomorphic MLPs, and the latter two have an output dimension of size $N_h$, which we set of $N_h=4$ in out simulations. We also include a term capturing the electron-electron Kato cusp conditions in Eq.~\eqref{eq:ee_cusp}.

\subsection{Variational wave function model in second quantization}
In second quantization, the anti-symmetry is encoded into the anti-commutation properties of the creation and annihilation operators. Therefore, no particle-permutation constraints have to be enforced, widening the range of functions to approximate the wave function~\cite{barrett2022autoregressive}, and we choose a holomorphic neural network: the Restricted Boltzmann Machine (RBM), which has successfully been applied to quantum chemistry applications~\cite{choo2020fermionic}. The time-dependent RBM for $N_h$ hidden parameters in the occupation number basis reads 
\begin{align}
    \Psi(n, t) &= \sum_{h_i \in \{-1, 1\}} e^{\sum_{i,j}  W_{ij}(t) h_i n_j + \sum_i b_i(t) h_i + \sum_j a_j(t) n_j} \\
    &= e^{\sum_j a_j(t) n_j} \prod_i 2 \cosh \left(b_i(t) + \sum_i W_{ij}(t) n_j\right)
\end{align}
where we introduced the spin-orbital index $j$, the time-dependent weights $W \in \mathbb{C}^{N_h \times 2M}$ and biases $b \in \mathbb{C}^{N_h}$ and $a \in \mathbb{C}^{2M}$ are taken to be complex.

\section{Monte Carlo sampling}
To obtain samples from the unnormalized Born probability density $\Psi(x)$, we use Markov-Chain Monte Carlo sampling. In first quantization these are obtained by random walks in configuration space with an update rule
\begin{align}
[\vec{r}_1, ..., \vec{r}_N] \to [\vec{r}_1, ..., \vec{r}_N] + \vec{u}
\end{align}
where $\vec{u} \sim \mathcal{N}(0, \epsilon^2) \in \mathbb{R}^{N \times D}$ where $\epsilon$ is chosen such that the Metropolis-Hastings acceptance rate is around $50 \%$. For molecules, we rely on Metropolis-Adjusted Langevin sampling, which uses the gradient of the wave function to improve the decorrelation between subsequent samples in the Markov chains~\cite{schatzle2023deepqmc}.
In second quantization, new samples are proposed by changing the spin-orbital occupied by a randomly chosen electron, while maintaining the number of electrons per spin sector. To reduce the correlations between the subsequent samples of a Markov Chain, we only keep every $N_T$'th accepted sample (referred to as a thinning factor). Hereby, $N_T = 15 \times d \times N$ to guarantee very low correlations.

\section{Scalability analysis}\label{sec:scaling}
We will discuss the scalability of our approach in terms of the number of parameters $N_p$ and the number of electrons $N$. We focus on models defined in continuous space and refer to Ref.~\cite{schmitt2020quantum}, for a similar analysis in a discrete basis set.

First, the number of parameters in a model scales as $\order{N \times M}$ for a mean field Slater determinant, and at least as $\order{1}$ for the Jastrow and backflow functions. 
Computing the wave function introduces a cost of $\order{N^3}$ due to the presence of the anti-symmetrizing Slater determinant in Eq.~\eqref{eq:model_general}. Computing the backflow and Jastrow functions involves combining pair-wise interactions between electrons, scaling at least as $\order{N^2}$ each.

To evolve a variational wave function in time using tVMC, we must compute the energy in Eq.~\eqref{eq:tvmc_energy}, the energy gradients in Eq.~\eqref{eq:tvmc_force}, and the QGT in \eqref{eq:tvmc_qgt}. The scaling of the computational cost with the number of electrons is dominated by computing the Laplacian in the local energy, which scales asymptotically as $\order{N^4}$~\cite{pfau2020ab, schatzle2023deepqmc}. We compute the Laplacian with the method recently introduced in Ref.~\cite{li_computational_2024}.

The cost of parameter gradients using automatic differentiation in Jax~\cite{jax2018github} scales the same as a forward pass. However, solving the TDVP equation~\ref{eq:tdvp} requires the inversion of the QGT, which costs $\order{N_p^3}$ and dominates the computational burden when the number of parameters increases. The latter can be reduced to $\order{N_p}$ in tre-tVMC using the minSR approach from Ref.~\cite{chen2024empowering}. 

\section{Orbital rotation}\label{sec:orbital_rotation}
In the Hartree-Fock (HF) molecular orbital basis, the ground state wave function is peaked around the basis state corresponding to the HF solution (we will denote this basis state with $x^{\textrm{HF}}$). This property often makes optimization challenging, as pointed out in Ref.~\cite{choo2020fermionic}.
The difficulty in the variational optimization follows from the energy gradients that tend to vanish for finite sample sets~\cite{sinibaldi2023unbiasing}. More specifically, for a low number of samples and near the ground state solution, samples generated from $\abs{\Psi(x)}^2$ will be dominated by $x^{\textrm{HF}}$. As a result, VMC will often optimize the variational state until it reaches the HF ground state. One effective solution to this problem is to perform a basis rotation and express the Hamiltonian in a basis that is more suitable for Monte Carlo estimates of the energy and its gradient. A flexible approach is to define a \emph{fixed} (i.e.\ non-variational) state $\Upsilon(x)$ for which we aim to minimize the energy. For example, we choose $\Upsilon(x) = \textrm{cte}$ to be the uniform superposition state at a fixed electron number. To obtain the minimal energy, we optimize the (real) single-particle orbital rotation matrix operator $U = \exp\left[ \sum_{\sigma, i, j} A_{ij} \hat{a}_{i,\sigma}^\dagger \hat{a}_{j,\sigma}\right]$, i.e.\ $\hat{\tilde{a}}_{i,\sigma}^\dagger = \sum_j O_{ij} \hat{a}_{i,\sigma}^\dagger$ with $O = \exp\left[A_{ij}\right]$. To guarantee the unitarity of $U$, in practice, we parameterize the upper-triangle matrix of a matrix $B$ and set $A = B - B^T$. Given the pre-computed one-body and two-body reduced density matrices
\begin{align}
    \Gamma^\sigma_{i,j}[\Upsilon] &= \left<\hat{a}_{i,\sigma}^\dagger \hat{a}_{j, \sigma}\right>_{\Upsilon} \\
    \Gamma^\sigma_{i,j,k,l}[\Upsilon] &= \left<\hat{a}_{i,\sigma}^\dagger \hat{a}_{j,\sigma'}^\dagger \hat{a}_{k, \sigma'} \hat{a}_{l, \sigma}\right>_{\Upsilon}
\end{align}
one can efficiently evaluate the energy and its gradients~\cite{moreno2023enhancing}. The result is a new basis $\hat{\tilde{a}}_{i,\sigma}^\dagger = \sum_j O_{ij} \hat{a}_{i,\sigma}^\dagger$ in which we represent our wave function model.

\section{Molecules: electric profile, ground states and integration error\label{sec:H2_gs}\label{sec:electric_profiles}}
\begin{figure}
    \centering
    \includegraphics[width=0.5\textwidth]{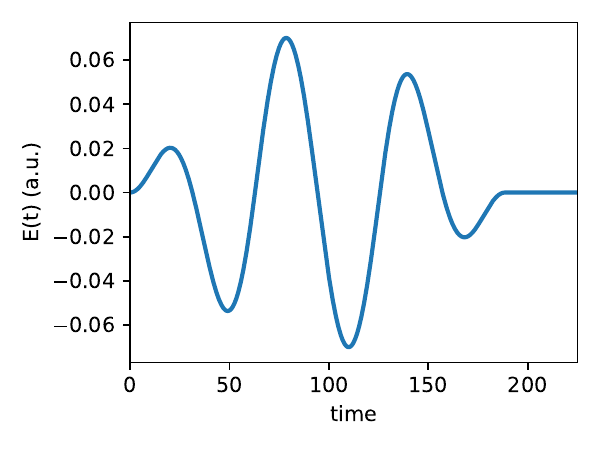}
    \caption{Time-dependent electric field applied to the $H_2$ molecule.}
    \label{fig:electric_field_H2}
\end{figure}

For the $H_2$ molecule, we use a laser profile of the form (see Fig.~\ref{fig:electric_field_H2})~\cite{li2005time}
\begin{align}
    \Epsilon(t) = \Epsilon^{\textrm{max}} \sin \left(\omega t\right) \times
    \begin{cases}
      0 & (t < 0) \\
      t/T & (0 \leq t < T) \\
      1 & (T \leq t < 2T) \\
      3-t/T & (2T \leq t < 3T)\\
      0 & (3T \leq t)
    \end{cases}      
\end{align}
where $T = 2\pi/\omega$. The corresponding time-dependent potential reads
\begin{align}
    V_{\textrm{ext}}(x, t) = - \sum_{i=1}^N\vec{\Epsilon}(t) \cdot \vec{r}_i
\end{align}
where $\vec{\Epsilon}(t) = \Epsilon(t) \vec{e}_z$, with $\vec{e}_z$ is the unit vector along the $z$-axis.

\begin{figure}[h!]
    \includegraphics[width=0.30\textwidth]{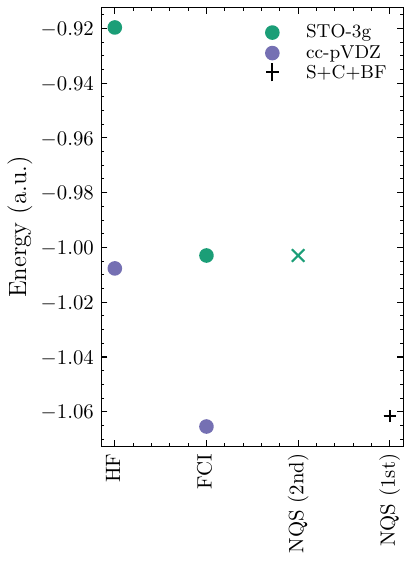}%
    \caption{Ground-state energies of $H_2$ at twice the equilibrium distance, obtained in second quantization with various atomic basis sets using HF, FCI and NQS (RBM), as well as in first quantization with NQS (S+C+BF, see Fig.~\ref{fig:H2_tevo}), as described in Section~\ref{sec:model_first_quant_molec}. \label{fig:H2_gs}}
\end{figure}

\begin{figure}[h!]
    \centering
    \includegraphics[width=0.5\textwidth]{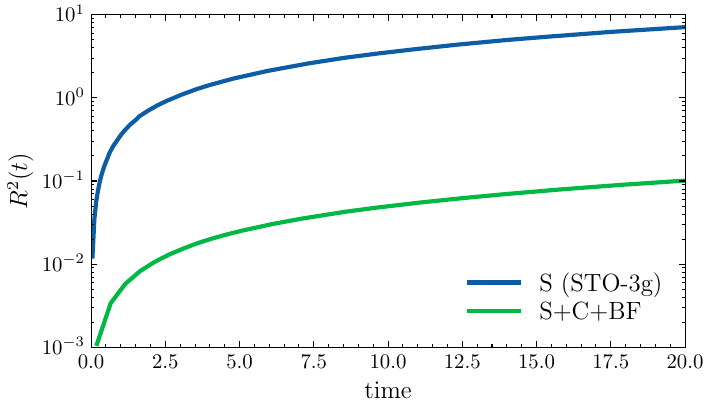}
    \caption{Integrated $R^2(t)$ as a function of time for the $H_2$ molecule in a laser field for the backflow model (S+C+BF), in comparison with a pure mean-field model with STO-3g orbitals (S).}
    \label{fig:integration_error_h2s}
\end{figure}

The real-time evolutions in this work start from the approximated ground state of some initial electronic Hamiltonian. We summarize the ground state energies obtained with VMC using the neural wave function models in Fig.~\ref{fig:H2_gs}, and compare to results obtained with HF and FCI. We also provide the integration error of the neural backflow model in Fig.~\ref{fig:integration_error_h2s} and compare to the integration error obtained with a pure mean-field approach in the STO-3g basis.

\section{Quantum dot: monopole}\label{sec:quantumdot_monopole}
\begin{figure}[htb]
    \centering
    \includegraphics[width=0.5\textwidth]{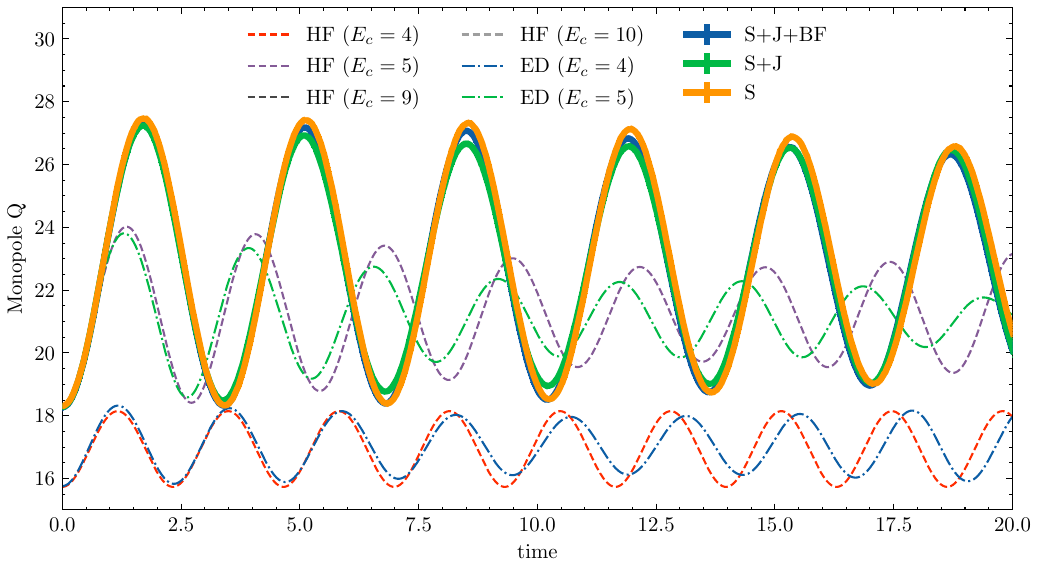}
    \caption{The electric monopole $Q$ as a function of time for a fully polarized quantum dot with $N=6$ and subject to a quench $\kappa(t) = 1 \to 2$ at $t=0$. We show predictions from TD-HF and ED in the Laguerre basis sets in Appendix~\ref{sec:model_first_quant_qd}, and compare with predictions using tVMC with a single Slater determinant (S), a Slater-Jastrow model (S+J), and a Slater-Jastrow-Backflow (S+J+BF) model.}
    \label{fig:quantumdot_monopole_1_2}
\end{figure}

Figure~\ref{fig:quantumdot_monopole_1_2} depicts the electric monopole for the quantum dot system in the main text. We observe the breathing modes of the electric monopole in Fig.~\ref{fig:quantumdot_monopole_1_2}. By comparing TD-HF and ED predictions, we conclude that the breathing behavior strongly depends on the chosen basis set (determined by the cutoff $E_c$, see Appendix~\ref{sec:model_first_quant_qd}), and whether correlations are accounted for. Using tVMC, we are able to capture the electronic correlations by introducing a time-dependent Jastrow factor and backflow transformations.

To obtain the TD-HF results in Fig.~\ref{fig:quantumdot_monopole_1_2}, we use the matrix elements in the Laguerre-Gauss basis in 2D in Eq.~\eqref{eq:laguerre_gauss_basis} from Ref.~\cite{anisimovas1998energy}.

\section{Pair correlation function}\label{sec:pair_correlation_derivation}
The first two terms in the integrated pair-correlation function in Eq.~\eqref{eq:integrated_G2} can be evaluated analytically (we again assume a fully polarized system so we can ignore spin):
\begin{align}
    T_1 &=\frac{1}{\mathcal{N}} \int \dd \vec{x} \dd \vec{y} \mel{\Psi}{\phi^\dagger (\vec{x}) \phi (\vec{x}) \phi^\dagger (\vec{y}) \phi (\vec{y})}{\Psi} \\
    &= \frac{1}{\mathcal{N}} \int \dd \vec{x} \dd \vec{y} \int \dd x' \abs{\Psi(x')}^2 \sum_{i,j}\delta(\vec{r}_i' - \vec{x})\delta(\vec{r}_j' - \vec{y}) \\
    &=\frac{N(N-1)}{\mathcal{N}}  \int \dd x' \abs{\Psi(x')}^2 \int \dd \vec{x} \dd \vec{y} \delta(\vec{r}_1' - \vec{x})\delta(\vec{r}_2' - \vec{y}) \\
    &=\frac{N(N-1)}{\mathcal{N}}  \int \dd x \abs{\Psi(x)}^2  \\
    &=N(N-1)  
\end{align}

Furthermore
\begin{align}
    T_2 &= \frac{1}{\mathcal{N}^2} \int \dd \vec{x} \dd \vec{y} \mel{\Psi}{\phi^\dagger (\vec{x}) \phi (\vec{x})}{\Psi} \mel{\Psi}{\phi^\dagger (\vec{y}) \phi (\vec{y})}{\Psi}   \\
    &= \frac{1}{\mathcal{N}^2} \left( \int \dd x' \abs{\Psi(x')}^2 \int \dd \vec{x} \sum_i \delta(\vec{r}_i' - \vec{x})  \right) \nonumber \\ 
    &\phantom{=} \times \left( \int \dd x'' \abs{\Psi(x'')}^2 \int \dd \vec{y} \sum_j \delta(\vec{r}_j'' - \vec{y})  \right) \\
    &= N^2
\end{align}

For the last term we have
\begin{align}
    T_3 &=\frac{1}{\mathcal{N}^2} \int \dd \vec{x} \dd \vec{y} \mel{\Psi}{\phi^\dagger (\vec{x}) \phi (\vec{y})}{\Psi} \mel{\Psi}{\phi^\dagger (\vec{y}) \phi (\vec{x})}{\Psi} \\
    &= \frac{1}{\mathcal{N}^2} \int \dd \vec{x} \dd \vec{y} \int \dd x' \dd x'' \Psi^*(x') \Psi(x'') \sum_i \delta(\vec{r}_i' - \vec{x}) \delta(\vec{r}_i'' - \vec{y}) \delta(x'_{\neg i} - x''_{\neg i}) \nonumber \\
    & \phantom{=} \times \int \dd x''' \dd x'''' \Psi^*(x''') \Psi(x'''') \sum_j \delta(\vec{r}_j''' - \vec{y}) \delta(\vec{r}_j'''' - \vec{x}) \delta(x'''_{\neg j} - x''''_{\neg j}) \\
    &= \frac{N^2}{\mathcal{N}^2} \int \dd x' \dd x''''  \Psi^*(x') \Psi(x'''')  \int \dd \vec{x} \delta(\vec{r}_1' - \vec{x}) \delta(\vec{r}_1'''' - \vec{x}) \nonumber \\
    &\phantom{=} \times \int \dd x'' \dd x'''  \Psi^*(x'') \Psi(x''')  \int \dd \vec{y} \delta(\vec{r}_1'' - \vec{y}) \delta(\vec{r}_1''' - \vec{y}) \nonumber \\
    &\phantom{=} \times \delta(x'_{\neg 1} - x''_{\neg 1}) \delta(x'''_{\neg 1} - x''''_{\neg 1})  \\
    &= \frac{N^2}{\mathcal{N}^2} \int \dd x' \dd x''' \Psi^*(x') \Psi(\vec{r}_1',  x_{\neg 1}''') \Psi^*(x''') \Psi(\vec{r}_1''',  x_{\neg 1}')   \nonumber \\
    &=\frac{N^2}{\mathcal{N}^2} \int \dd x' \dd x''' \abs{\Psi(x')}^2 \frac{\Psi(\vec{r}_1',  x_{\neg 1}''')}{\Psi(x')} \abs{\Psi(x''')}^2 \frac{\Psi(\vec{r}_1''',  x_{\neg 1}')}{\Psi(x''')} 
\end{align}
where we introduced the notation $x_{\neg i} = \left[\vec{r}_1 ,..., \vec{r}_{i-1}, \vec{r}_{i+1},...,\vec{r}_N\right]$. In going from the second to the third line we used the following identity
\begin{align}
    \int \dd x'''' \Psi(x'''') \delta(x'''_{\neg 1} - x''''_{\neg 1})  \int \dd \vec{x} \delta(\vec{r}_1' - \vec{x}) \delta(\vec{r}_1'''' - \vec{x})  &= \Psi(\vec{r}_1', x'''_{\neg 1})
\end{align}
Hence we obtain the Monte Carlo estimator
\begin{align}
    G^{(2)} = &\int \dd \vec{x} \dd \vec{y}  g^{(2)}(\vec{x}, \vec{y}) \\
    =& N \left( N \mathbb{E}_{ \substack{x\sim \abs{\Psi}^2  \\ x'\sim \abs{\Psi}^2}} \left[ \frac{\Psi(\vec{r}_1', \vec{r}_2,...,\vec{r}_N)}{\Psi(\vec{r}_1, \vec{r}_2,...,\vec{r}_N)} \frac{\Psi(\vec{r}_1, \vec{r}_2',...,\vec{r}_N')}{\Psi(\vec{r}_1', \vec{r}_2',...,\vec{r}_N')}\right] - 1\right)
\end{align}
As an illustration, assume two particles ($N=2$) in a mean-field Slater determinant state. For simplicity of notation, we assume orthonormal and real orbitals $\{A(\vec{r}), B(\vec{r})\}$ and $\mathcal{N} = 1$, but the results hold in general. We have that
\begin{align}
    & \int \dd x \dd x' \Psi^*(x) \Psi(\vec{r}_1,  x_{\neg 1}') \Psi^*(x') \Psi(\vec{r}_1',  x_{\neg 1}) \\
    &= \int \dd \vec{r}_1 \dd \vec{r}_2\dd \vec{r}_1' \dd \vec{r}_2' \Psi^*(\vec{r}_1, \vec{r}_2) \Psi(\vec{r}_1,  \vec{r}_2') \Psi^*(\vec{r}_1', \vec{r}_2') \Psi(\vec{r}_1',  \vec{r}_2)
\end{align}
Using the orthonormality of the orbitals $A$ and $B$ we get
\begin{align}
    &\int \dd \vec{r}_2' \Psi(\vec{r}_1,  \vec{r}_2') \Psi(\vec{r}_1', \vec{r}_2') \\
    &= \int \dd \vec{r}_2 \Psi(\vec{r}_1,  \vec{r}_2) \Psi(\vec{r}_1', \vec{r}_2)\\
    &= \frac{1}{N} \left[A(\vec{r}_1) A(\vec{r}_1') + B(\vec{r}_1) B(\vec{r}_1')\right] 
\end{align}
such that (again using the orthonormality of the orbitals)
\begin{align}
    &\int \dd x \dd x' \Psi^*(x) \Psi(\vec{r}_1,  x_{\neg 1}') \Psi^*(x') \Psi(\vec{r}_1',  x_{\neg 1}) \\
    &= \frac{1}{N^2}\int \dd \vec{r}_1 \dd \vec{r}_1' \left[A(\vec{r}_1)^2 A(\vec{r}_1')^2 + B(\vec{r}_1)^2 B(\vec{r}_1')^2 + 2 A(\vec{r}_1) B(\vec{r}_1) A(\vec{r}_1')B(\vec{r}_1')\right]  \\
    &= \frac{1}{N^2} N = \frac{1}{N}
\end{align}
Altogether, we find
\begin{align}
    G^{(2)}_{MF} &= N(N-1) - N^2 + N^2 \frac{1}{N} = 0
\end{align}
as expected.

In the case of two spin types $\sigma=\{\uparrow, \downarrow\}$, $\vec{x}$ and $\vec{y}$ are also associated with spin quantum numbers $\sigma_{\vec{x}/\vec{y}} \in \{\uparrow, \downarrow\}$. To obtain $G^{(2)}$ we also sum over all possible spin projections
\begin{align}
    G^{(2)} =& \sum_{\sigma_{\vec{x}}, \sigma_{\vec{y}} \in \{\uparrow, \downarrow\}}\int \dd \vec{x} \dd \vec{y}  g^{(2)}(\vec{x}, \vec{y}). \label{eq:integrated_G2_spin}
\end{align}
Notice, however, that the Monte Carlo estimators above change when spin is accounted for. We can distinguish two cases: $ \sigma_{\vec{x}} = \sigma_{\vec{y}} (=\sigma)$ (denoted $\parallel$) and $\sigma_{\vec{x}} \neq \sigma_{\vec{y}}$ (denoted $\perp$) we have
\begin{align}
    T_1^\parallel &= N_\sigma (N_\sigma-1)\\
    T_1^\perp &= N_\uparrow N_\downarrow 
\end{align}
and 
\begin{align}
    T_2^\parallel &= N_\sigma^2\\
    T_2^\perp &= N_\uparrow N_\downarrow 
\end{align}
and lastly ($T_3^\parallel$ is again similar to $T_3$),
\begin{align}
    T_3^\perp &= 0
\end{align}

\section{Time-dependent harmonic interaction model}\label{sec:him}
We consider the time-dependent Hamiltonian in 1D
\begin{align}
    V(x, t<0) = \frac{\omega_0^2}{2}\sum_i \vec{r}_i^2 + g \sum_{i<j} (\vec{r}_i - \vec{r}_j)^2 \label{eq:HI_V_prequench} \\
    V(x, t\geq 0) = \frac{\omega_f^2}{2}\sum_i \vec{r}_i^2 + g(t) \sum_{i<j} (\vec{r}_i - \vec{r}_j)^2
\end{align}
with $g(t)$ to be determined later. We quench the confining harmonic potential $\omega(t): \omega_0 \to \omega_f$ at $t=0$.
We can use the scaling approach to quantum non-equilibrium dynamics of many-body systems from Ref.~\cite{gritsev2010scaling} to obtain the time-dependent wave function and interaction strength $g(t)$. We start from the ground state of the Hamiltonian at $t<0$: $\Psi_0(x)$. 
The ground state with trap frequency $\omega_0$ is given by
\begin{align}
    \Phi(x) = \det \mathcal{V}(x) e^{-\frac{\alpha}{2} \sum_i r_i^2-\frac{\beta}{2} (\sum_i \vec{r}_i)^2} 
\end{align}
with $\alpha = \gamma$, and $\beta=(1-\gamma)/N$, where $\gamma = \sqrt{1+N g}$ and $\det\mathcal{V}$ the Vandermonde determinant, i.e.\ $\det\mathcal{V}(x) = \prod_{i<j} (\vec{r}_i - \vec{r}_j)$.

The solution to the TDSE can then be written (up to a time-dependent overall phase factor)
\begin{align}
    \Psi(x, t) = e^{-i\frac{F(t)}{2}\sum_i x_i^2} \Phi\left(\frac{x}{L(t)}, \tau(t)\right)
\end{align}
with $\Phi$ the solution to the time-independent Hamiltonian with trap frequency $\omega_0$ and $\tau(t) = \int dt L^{-2}(t)$.
By solving the Ermakov equation, we obtain the scaling (which describes the breathing-mode behavior)
\begin{align}
    L(t) = \sqrt{A \cos (2\omega_f t)+C} 
\end{align}
with constants
\begin{align}
    A = \frac{\omega_f^2 - \omega_0^2}{2\omega_f^2} \\
    C = \frac{\omega_f^2 + \omega_0^2}{2\omega_f^2}
\end{align}
Note that $A \cos (2\omega_f t)+C \geq 0$ since $C-A = (\omega_0/\omega_f)^2$. We then have
\begin{align}
    F(t) &= -\frac{\dot{L}(t)}{L(t)} \\
    &= A \omega_f  \frac{\sin (2 \omega_f t)}{L(t)^2}
\end{align}
such that at $F(t=0)=0$. Furthermore
\begin{align}
    \tau(t) = \frac{1}{\omega_0} \arctan \left(\frac{\omega_0}{\omega_f} \tan (\omega_f t)\right)
\end{align}
such that at $t=0$ we have $\tau = 0$.

\section{Details of the tre-tVMC method}\label{sec:tretvmc_discussion}
A key component of the tre-tVMC method is the estimation of the fidelity with Monte Carlo sampling. Due to the non-unitarity of the root operators $\hat{R}_k$ the norm change reflected by the denominator of the fidelity $\mel{\Psi(t)}{\hat{R}_k^\dagger\hat{R}_k}{\Psi(t)}$ becomes nontrivial. Estimating this factor requires the evaluation of higher-order powers of $\hat{H}$ which we aim to avoid by design. 
To bypass these technical difficulties, we introduce a novel Monte-Carlo estimator of the fidelity, by relying on self-normalized importance sampling~\cite{nys2024real}
\begin{align}
\mathcal{F}\left(\hat{R}\ket{\Psi},\ket{\Phi(\theta)}\right) &=  \mathcal{N} \frac{\mel{\Psi}{\hat{R}^\dagger}{\Phi(\theta)} \mel{\Phi(\theta)}{\hat{R}}{\Psi} }{\braket{\Psi}\braket{\Phi(\theta)}}, \label{eq:Rfidelity} \\
&= \frac{\mathbb{E}_{(x, y) \sim \chi} \left[ w(y) F_\textrm{loc}(x, y) \right] }{\mathbb{E}_{(x,y)\sim \chi} \left[ w(y) \right]} \label{eq:fidelity_weighted}.
\end{align}
where we introduced the norm ratio $\mathcal{N} = \mel{\Psi}{\hat{R}^\dagger\hat{R}}{\Psi}/\braket{\Psi}$, the joint probability density $\chi(x, y;\theta) = \smash{\abs{\Phi(x;\theta)}^2 \abs{\Psi(y)}^2}$ and local fidelity estimator~\cite{sinibaldi2023unbiasing} and weight
\begin{align}
    F_\textrm{loc}(x, y) &= \tfrac{ \left[\hat{R}\Psi\right](x) }{\Phi(x;\theta)} \tfrac{\Phi(y; \theta)}{ \left[\hat{R}\Psi\right](y) },\\
    w(y) &= \abs{\tfrac{ \left[\hat{R}\Psi\right](y) }{\Psi(y)} }^2. 
\end{align}
In addition, we also use the Control-Variates approach introduced in Ref.~\cite{sinibaldi2023unbiasing} : $F_\textrm{loc}(x, y) \to F_\textrm{loc}(x, y) - \tfrac{1}{2} \left(\smash{\abs{F_\textrm{loc}(x, y)}^2 - 1}\right)$, to improve the signal-to-noise ratio in the fidelity estimator.

Finally, we discuss some of the underlying choices made in the derivation of our method, and some of the beneficial properties of its final form. 
As a variation of our method, the product form in Eq.~\eqref{eq:prod} can be chosen to contain more than $K$ terms, thereby allowing additional flexibility of the solutions for the reduced time steps $c_k \dt$ at the cost of additional optimizations per time step. 
Lastly, other forms of the MC fidelity estimator are possible. Our choice of the latter is specifically designed with the following stringent requirements. We aim to avoid sampling from operator states $\hat{R}_k\ket{\Psi(t)}$, since this would introduce an additional problematic scaling with the number of particles in the sampling MC procedure. Furthermore, it would require many evaluations of the kinetic operator in continuous space, which is expensive~\cite{schatzle2023deepqmc}. Hence, while such an approach would not require us to separately estimate the norm ratio $\mathcal{N}$ in Eq.~\eqref{eq:fidelity_weighted}, it would introduce a significant computational overhead. This approach would, however, need to be considered when considering the implicit midpoint rule~\cite{gutierrez2022real} together with fidelity optimization~\cite{sinibaldi2023unbiasing}. Nevertheless, our MC estimator is chosen such that the $\mathcal{N}$ ratio has an irrelevant and negligible effect on the gradients of the fidelity: since $\mathcal{N}$ is parameter independent and enters as an overall factor, it does not affect the direction of the fidelity gradients. Lastly, by recasting the MC fidelity in the form of Eq.~\eqref{eq:fidelity_weighted}, we avoid the need for computing parameter gradients on the operator state $\hat{R}_k^\dagger \ket{\Phi(\theta)}$, thereby avoiding the need to compute gradients of local operators.

\begin{figure}[tbh]
    \centering
    \includegraphics[width=0.35\linewidth]{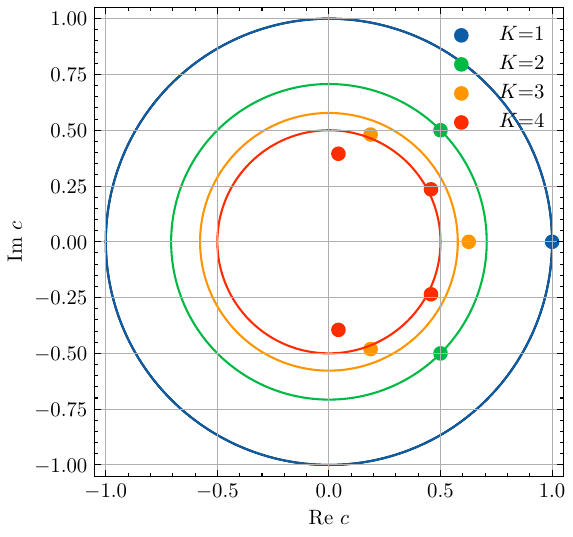}
    \includegraphics[width=0.30\linewidth]{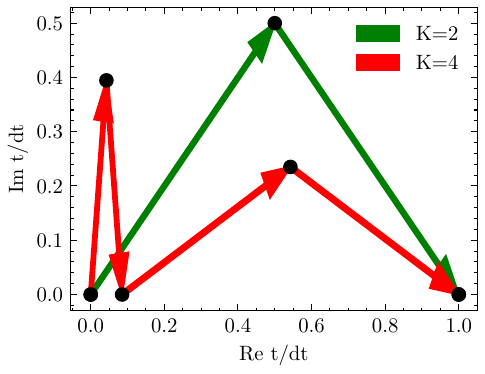}
    \caption{(left) Solutions to the tre-tVMC equations in Eq.~\eqref{eq:prod} for $K=1,2,3,4$. The circles depict $1/\sqrt{K}$ for reference. (right) Integration path in the tre-tVMC approach for a time step $\dt$ and $K=2, 4$. The black dots show $C(l) = dt\sum_{k=1}^l c_k$ according to the chosen orderings.}
    \label{fig:circle_hop}
\end{figure}

We visualize the $c_k$ solutions to the tre-tVMC equations in Figure~\ref{fig:circle_hop}. For example, in our experiments we use $K=4$, where $c_{1,2} \approx 0.0426 \pm 0.3946i$, $c_{3,4} \approx 0.4574 \pm 0.2351 i$. We observe that the $c_k$ solutions scale approximately as $\abs{c_k} = 1/\sqrt{K}$. This allows us to increase the time-step for higher orders $K$. The chosen integration paths are depicted in the right panel of Fig.~\ref{fig:circle_hop} for a single time step.

In principle the fidelity optimization can be carried out with standard optimization techniques such as Adam. However, a major speedup and increase in optimization stability is obtained by updating the fidelity gradients with Stochastic Reconfiguration and a \emph{block-diagonal approximation to the quantum geometric tensor}, where blocks correspond to layers in the neural network. The block structure depends on the form of the model and only considers inter-layer correlations. More importantly, by avoiding the estimation of said correlation, we observe that the optimization is more stable, and one can carry out the time evolution even with a low number of samples.

\section{Numerical details}\label{sec:numerical_details}
For tVMC, we use around $500k$ samples to obtain accurate estimates of the forces and QGT. The TDVP equations are solved using the smooth SVD technique from Ref.~\cite{medvidovic2024neural}, using a cutoff in the range $\texttt{acond} \in [10^{-8}, 10^{-3}]$, depending on the singular-value spectrum of the ground state QGT. For tVMC, we use a thinning factor $N_T = 5 N d$ to guarantee complete decorrelation between the samples and use an adaptive Gaussian random walk (or Metropolis-Adapted Langevin sampler~\cite{vicentini2022netket, schatzle2023deepqmc} for the molecular case) sampler that targets $50-60 \%$ accuracy. We use an explicit Runge-Kutta method of order $3(2)$ and determine the time step based on the integration error.

For tre-tVMC, we use $18-36k$ samples (though less are possible) to estimate the fidelity and its gradients, and typically reduce the thinning factor. We take a learning rate of $0.03$ and $60$ optimization steps. The time step is determined based on the integration error $r^2$ and on energy conservation in dynamics subject to time-independent Hamiltonians.

\end{document}